\documentclass[useAMS,usenatbib]{mn2e}
\usepackage[utf8]{inputenc}
\usepackage[usenames,dvipsnames]{color}
\usepackage{graphicx}
\usepackage{hyperref}
\usepackage{amsmath}
\usepackage{amssymb}
\usepackage{fancyhdr}
\usepackage{cancel}
\usepackage{txfonts}
\usepackage[T1]{fontenc}
\usepackage{ae,aecompl}

\newcommand{\Sref}[1]{Section \ref{#1}}
\newcommand{\Tref}[1]{Table \ref{#1}}
\sfcode`\.=1001\sfcode`\?=1001\sfcode`\!=1001

\newcommand{\Fref}[1]{\ifhmode \ifnum\spacefactor=1001 Figure \ref{#1}\else Fig.\ \ref{#1}\fi \else Figure \ref{#1}\fi}
\newcommand{\Eref}[1]{\ifhmode \ifnum\spacefactor=1001 Equation (\ref{#1})\else equation (\ref{#1})\fi \else Equation (\ref{#1})\fi}

\newcommand{\ms}{\ensuremath{\textrm{m\,s}^{-1}}}
\newcommand{\kms}{\ensuremath{\textrm{km\,s}^{-1}}}
\newcommand{\SN}{\ensuremath{\textrm{S/N}}}
\newcommand{\chisq}{\ensuremath{\chi^2}}
\newcommand{\chisqn}{\ensuremath{\chi^2_\nu}}

\newcommand{\zem}{\ensuremath{z_\text{\scriptsize em}}}
\newcommand{\zab}{\ensuremath{z_\text{\scriptsize abs}}}

\newcommand{\ion}[2]{\ensuremath{\textrm{#1\,{\scshape{#2}}}}}
\newcommand{\tran}[3]{\ensuremath{\ion{#1}{#2}\,\lambda\textrm{#3}}}
\newcommand{\varal}{\ensuremath{\Delta \alpha/\alpha}}
\newcommand{\angstrom}{\mbox{\normalfont\AA}}

\usepackage{natbib}
\bibpunct{(}{)}{;}{a}{}{,}

\def \aap{A\&A}

\def \apss{Ap\&SS}

\def \apj{ApJ}

\def \memsai{Mem.~Soc.~Astron.~Italiana}
\def \mnras{MNRAS}

\title[Stringent limit on variation in $\alpha$]{High-precision limit on variation in the fine-structure constant from a single quasar absorption system}

\author[S. M. Kotuš, M. T. Murphy and R. F. Carswell]{S. M. Kotuš$^{1}$\thanks{E-mail: skotus@swin.edu.au (SMK)}, M. T. Murphy$^{1}$\thanks{E-mail: mmurphy@swin.edu.au (MTM)}  and R. F. Carswell$^{2}$\\
$^{1}$Centre for Astrophysics and Supercomputing, Swinburne University of Technology, Hawthorn, Victoria 3122, Australia\\
$^{2}$Institute of Astronomy, University of Cambridge, Madingley Road, Cambridge CB3 0HA, UK}

\begin{document}

\date{Accepted 2016 Sep 08. Received 2016 Aug 16; in original form 2016 March 20}

\pagerange{\pageref{firstpage}--\pageref{lastpage}} \pubyear{2016}

\maketitle

\label{firstpage}

\begin{abstract}
The brightest southern quasar above redshift $z=1$, HE 0515$-$4414, with its strong intervening metal absorption-line system at $\zab=1.1508$, provides a unique opportunity to precisely measure or limit relative variations in the fine-structure constant ($\varal$). A variation of just $\sim$3\,parts per million (ppm) would produce detectable velocity shifts between its many strong metal transitions. Using new and archival observations from the Ultraviolet and Visual Echelle Spectrograph (UVES) we obtain an extremely high signal-to-noise ratio spectrum (peaking at $\SN\approx250$\,pix$^{-1}$). This provides the most precise measurement of $\varal$ from a single absorption system to date, $\varal=-1.42\pm0.55_{\rm stat}\pm0.65_{\rm sys}$\,ppm, comparable with the precision from previous, large samples of $\sim$150 absorbers. The largest systematic error in all (but one) previous similar measurements, including the large samples, was long-range distortions in the wavelength calibration. These would add a $\sim$2\,ppm systematic error to our measurement and up to $\sim$10\,ppm to other measurements using Mg and Fe transitions. However, we corrected the UVES spectra using well-calibrated spectra of the same quasar from the High Accuracy Radial velocity Planet Searcher (HARPS), leaving a residual 0.59\,ppm systematic uncertainty, the largest contribution to our total systematic error. A similar approach, using short observations on future, well-calibrated spectrographs to correct existing, high $\SN$ spectra, would efficiently enable a large sample of reliable $\varal$ measurements. The high $\SN$ UVES spectrum also provides insights into analysis difficulties, detector artifacts and systematic errors likely to arise from 25--40-m telescopes.
\end{abstract}


\section{Introduction}
The Standard Model of particle physics is incomplete because it cannot explain the values of fundamental constants, or predict their dependance on parameters such as time and space. Therefore, without a theory that is able to properly explain these numbers, their constancy can only be probed by measuring them in different places, times and conditions. Furthermore, many theories which attempt to unify gravity with the other three forces of nature invoke fundamental constants that are varying \citep[see][]{2011LRR....14....2U}.

In this work we focus on constraining the variability of the fine-structure constant, $\alpha \equiv e^2/ \hbar c$, which represents the coupling strength of the electromagnetic force. In the last 15 years there have been many attempts to measure this constant in absorption systems along the lines of sight to distant quasars. The approach called the ``Many Multiplet" (MM) method, pioneered by \citet{1999PhRvL..82..888D} and \citet{1999PhRvL..82..884W}, compares the relative velocity spacing between different metal ion transitions and relates it to possible variation in $\alpha$. For example, considering just a single transition, variation in $\alpha$ is related to the velocity shift $\Delta v_i$ of a transition

\begin{equation}
\varal\equiv\frac{\alpha_{\rm obs}-\alpha_{\rm lab}}{\alpha_{\rm lab}}\approx\frac{-\Delta v_i}{2c}\frac{\omega_i}{q_i}\,,
\end{equation}
where $c$ is the speed of light, $q_i$ is sensitivity of the transition to $\alpha$ variation, calculated from many body relativistic corrections to the energy levels of ions and $\omega_i$ is its wavenumber measured in the laboratory.
There have been two MM method studies of large absorber samples: the Keck High Resolution Echelle Spectrometer (HIRES/Keck) sample of 143 absorption systems \citep{1999PhRvL..82..884W, 2001PhRvL..87i1301W, 2001MNRAS.327.1208M, 2003MNRAS.345..609M, 2004LNP...648..131M} and the Ultraviolet and Visual Echelle Spectrograph on the Very Large Telescope (UVES/VLT) sample of 154 absorption systems \citep{2011PhRvL.107s1101W, 2012MNRAS.422.3370K}. These studies reported weighted mean values of $\varal=-5.7\pm1.1$\,parts per million (ppm) and $2.29\pm0.95$\,ppm, respectively. \citet{2011PhRvL.107s1101W} and \citet{2012MNRAS.422.3370K} also combined both samples and found a $4.1\sigma$ statistical preference for a dipole-like variation in $\alpha$ across the sky. Contrary to these studies, several other attempts to measure $\alpha$ from individual absorbers \citep[][]{2004A&A...415L...7Q, 2006A&A...451...45C, 2005A&A...434..827L, 2006A&A...449..879L, 2007A&A...466.1077L} or smaller samples of absorption systems \citep[][]{2004PhRvL..92l1302S, 2004A&A...417..853C} reported $\varal$ consistent with zero. However, these analyses contained shortcomings that produced biased values and/or considerably underestimated error bars (\citeauthor{2007PhRvL..99w9001M}, \citeyear{2007PhRvL..99w9001M}, \citeyear{2008MNRAS.384.1053M}, cf. \citeauthor{2007PhRvL..99w9002S}, \citeyear{2007PhRvL..99w9002S}). For example, a reanalysis including remodelling of the \citet[][]{2004A&A...417..853C} spectra by \citet{2015MNRAS.454.3082W}, yielded a weighted mean of $\varal = 2.2\pm2.3$\,ppm which, given the distribution of the 18 quasars across the sky, was not inconsistent with the evidence for dipole-like $\alpha$ variation.

The crucial observational question is whether a systematic effect, or some combination of multiple systematic effects, could mimic the results from the two large studies. Although early studies \citep{2001MNRAS.327.1223M, 2003Ap&SS.283..577M} did not report systematic effects that could strongly affect measurements of $\varal$ obtained from the large statistical samples, more recent analyses \citep[e.g.][]{2013MNRAS.435..861R} identified long-range distortions of the wavelength scale established from traditional ThAr calibration lamp spectra. These distortions were discovered via ``supercalibration" techniques in which solar or solar-like spectra from asteroids or ``solar twin" stars were used to establish an alternative wavelength scale. Earlier, similar supercalibration checks by \citet{2008A&A...481..559M}, \citet{2010ApJ...708..158G} and \citet{2010ApJ...723...89W}  did not reveal such long-range distortions. However, such systematic effects were confirmed by \citet{2015MNRAS.447..446W} who analysed 2 decades of archival solar twin spectra from UVES and HIRES. They found that long-range distortions were common and provided a compelling explanation for the non-zero values of $\varal$ found in the large samples of quasar spectra from Keck and VLT, particularly the latter.

From ``The UVES Large Program for testing Fundamental Physics", which was specifically designed to measure $\varal$, several recent measurements have been made. \citet{2013A&A...555A..68M} provided the tightest constraint on $\varal$ from an individual absorber, with a statistical precision of 2.4\,ppm and a systematic error estimated to be $\pm$1\,ppm. However, long-range distortions were not corrected in this study. Recently, \citet{2014MNRAS.445..128E} made 9 measurements of $\varal$ in 3 absorbers using 3 different telescopes, and used supercalibrations to correct for the long-range distortions. Their combined result indicated no variation in $\alpha$ at the $\approx$4\,ppm precision level, including estimates of any remaining systematic effects. This represents the only distortion-corrected measurement of $\varal$ so far. Therefore, there is a clear need to both make new, reliable measurements of $\varal$ and to further explore potential systematic effects. This is particularly important in view of the higher-quality spectra that will soon be available from better-calibrated spectrographs and larger telescopes.

In this paper we report the most robust and precise constraint on $\alpha$-variation in the well-studied $\zab=1.1508$ absorption system towards QSO HE 0515$-$4414. This is the brightest quasar at redshift above $z=1$ in the southern sky, so it offers the best opportunity to minimize statistical errors and study systematic effects beyond those discernible in lower-fidelity spectra. So far, three constraints on $\varal$ have been attempted in this absorber \citep[][hereafter Q04, L06, C06]{2004A&A...415L...7Q, 2006A&A...449..879L, 2006A&A...451...45C}. The L06 result was revised in \citet[][hereafter M08b]{2008EPJST.163..173M}. We discuss the results from these studies in the context of our new results in \Sref{comp_pre_mea}, suffice it to say here that there are several motivations to measure $\varal$ in this absorption system again.

Firstly, significantly more observational spectra are now available, including 3 times the number of UVES exposures previously used. \Tref{obs1} summarises all the available spectra obtained with the UVES spectrograph used in this study. When all the UVES spectra are combined, it results in the highest $\SN$ spectrum taken with a high-resolution optical spectrograph of a quasar, to our knowledge. This allows a thorough investigation of systematic effects which provides an insight into the problems likely to be faced when similar high-quality spectra are obtained from upcoming new spectrographs and telescopes.

\newcommand\oldtabcolsep{\tabcolsep}
\setlength{\tabcolsep}{0.5em}
\begin{table*}
\caption{Summary of the observations from UVES/VLT used in this work. `Project ID' represents the internal ESO project number in which exposures were taken. `Exposure time' represents the total observing time in each project, divided into dichroic (when both arms of the spectrograph were in use) and single-arm observations. The blue and red \SN\ correspond to neighbouring continuum to the \ion{Fe}{ii}\,1608 and \ion{Mg}{ii} doublet transitions, respectively. The seeing is the average during the project, with its range in parentheses. Years 1999a and 1999b correspond to 1$\times$2 and 1$\times$1 CCD binning, respectively. In year 1999a, the exposures in the 346-nm setting were taken in the blue arm only, while the 437-nm and 860-nm exposures were taken together in dichroic mode.}\label{obs1}
\centering
  \begin{tabular}{@{}l c c c c c c c c c c c c c c c }   
  \hline
  Project ID/Year       & \multicolumn{5}{c}{Number of exposures}            &  \multicolumn{2}{c}{Exposure time [s]}      &  \multicolumn{2}{c}{\SN}        &  \multicolumn{5}{c}{Slit width ["]}                          & Seeing ["]             \\
                        & 346     & 437   & 520     & 580    & 860           & Dichroic   & Single-arm                     & Blue & Red                       & 346        & 437       & 520     & 580        & 860          &                        \\
  \hline
  60.A-9022(A)/1999a    & $3$     & $2$   & $0$     & $0$    & $2$           & $9500$     & $14200$                        &  80  &                           & $0.8$   & $0.8$      &           &            & $0.7$        & $0.86$  $(0.6$--$1.4)$ \\
  60.A-9022(A)/1999b    & $0$     & $0$   & $1$     & $3$    & $0$           &            & $16600$                        &      &  103                      &         &            & $0.7$     & $0.7$      &              & $0.61$  $(0.5$--$0.8)$ \\
  66.A-0212(A)/2000     & $7$     & $6$   & $0$     & $4$    & $9$           & $53100$    &                                &  70  &  121                      & $0.8$   & $0.8$      &           & $0.8$      & $0.8$        & $0.61$  $(0.4$--$0.8)$ \\ 
  072.A-0100(A)/2003    & $0$     & $0$   & $0$     & $2$    & $0$           &            & $6120$                         &      &  68                       &         &            &           & $0.7$      &              & $0.81$  $(0.8$--$0.9)$ \\
  079.A-0404(A)/2007    & $3$     & $0$   & $0$     & $3$    & $0$           & $9900$     &                                &  22  &  56                       & $0.7$   &            &           & $0.7$      &              & $1.77$  $(1.7$--$1.8)$ \\
  082.A-0078(A)/0809    & $16$    & $0$   & $0$     & $16$   & $0$           & $46400$    &                                &  79  &  158                      & $0.6$   &            &           & $0.5$      &              & $1.00$  $(0.6$--$1.7)$ \\
  \hline 
  \end{tabular}
  \end{table*}
\setlength{\tabcolsep}{\oldtabcolsep}

Secondly, the previous work on this absorber has used only 6 \ion{Fe}{ii} transitions to constrain $\varal$, of which only one transition, $\lambda1608$, has a velocity shift in the opposite direction from the other transitions if $\alpha$ varies. If there exists a systematic error (e.g.~a calibration error) that causes the $\lambda1608$ line to shift, it would be seen as a non-zero $\varal$. Artificial shifts between this and other \ion{Fe}{ii} transitions are possible, particularly because it falls in a different arm of the UVES spectrograph in this absorption system. Artificial velocity shifts may also be due to the aforementioned long-range distortions. Therefore, in this work we include all transitions identified by \citet[][]{2014MNRAS.438..388M} that are useful for measuring $\varal$, which fall in the wavelength region of our spectra and which are not blended with absorbers from different redshifts. This somewhat increases the information available to reduce the statistical error on $\varal$, with the main motivation being the possibility of greater resistance to systematic effects.

Thirdly, it is apparent that the absorption profile models used in previous work significantly `under-fitted' the spectra, with 36 or fewer velocity components. According to \citet[][figure 8]{2008MNRAS.384.1053M}, under-fitting the spectra may lead to a substantial systematic error in the measured $\varal$ value. For this reason, we try to avoid this problem in our model by fitting as much of the statistically significant velocity structure as possible. This implies using 106 velocity components -- see \Sref{fitt} -- which is much larger than used in any previous MM analysis. We explore the possibility that we may still be under-fitting or somewhat `over-fitting' the spectra in \Sref{systematic_err}.

Finally, we use complementary observations from the High Accuracy Radial velocity Planet Searcher (HARPS) on the ESO 3.6 m La Silla telescope of the same object to recalibrate the wavelength scales of the UVES spectra. The supercalibration studies of \citet{2011A&A...525A..74M} and \citet{2015MNRAS.447..446W} demonstrated that the HARPS wavelength scale has much smaller (if any) long-range distortions than UVES and HIRES. By directly comparing the HARPS and UVES spectra we effectively transfer the relatively accurate HARPS wavelength scale onto the UVES spectra -- see \Sref{DC_method}. Additionally, we use the HARPS and complimentary Bench-mounted High Resolution Optical Spectrograph (bHROS) on the Gemini South Telescope, of very high resolution, $R\sim140000$, but with lower signal-to-noise ratio ($\SN$), to search for extra velocity structure (\Sref{unresolved}). Complementary observations from the HARPS and bHROS are described in Sections 2.2 and 2.3, respectively.

\section[]{Observations, data reduction and calibration}

HE 0515$-$4414 is a bright quasar with $V\approx 14.9$\,mag and redshift $\zem=1.71$ which was first identified in \citet[][]{1998A&A...334...96R}. We have used a large set of publicly available observations taken with three instruments: the UVES/VLT, HARPS/ESO$-$3.6\,m and bHROS/Gemini. We also publish the reduced spectra in \citet{srdan_kotus_2016_51715}. 

\subsection{UVES/VLT} \label{UVES_ORC}

The largest data set comprises 90 separate exposures from UVES/VLT \citep{2000SPIE.4008..534D}, which were observed between 1999 and 2009 in five different projects. The main $\varal$ constraint is from this dataset because it collectively has much higher $\SN$ ratio (per \kms) than the other spectra. In all of the UVES exposures, the slit was at the paralactic angle, projected perpendicular to the horizon. Three exposures in 2003 were taken with an iodine cell in the light path \citep[][]{2010ApJ...723...89W} and have significantly lower $\SN$ ratio and the expected forest of $\textrm{I}_{2}$ absorption features; they have been excluded from this analysis. Ten exposures do not have ``attached" ThAr exposures, which means that the echelle or cross-disperser gratings were likely reset between the quasar and ThAr exposures. This can cause velocity shifts and possibly distortions in the wavelength scale of a quasar spectrum and so we exclude these exposures as well. After these selections, the total number of useful UVES exposures reduces to 77 which have attached calibrations for measuring $\varal$ (\Tref{obs1}). These 77 exposures can also have velocity shifts and distortions in their wavelength scales, which we will account for in \Sref{DC_method}. Most of the exposures were taken by observatory staff (i.e.~``service mode"), except the three exposures taken in 2007, which were taken by visiting astronomers (i.e.~``visitor mode"). The exposure time of individual exposures is in the range between 2700 and 5400 s with a mean of $\sim$1\,h. The continuum around the bluest transition that we use in our study, \ion{Fe}{ii}\,1608 which falls near 3460\,\AA\ in the 346-nm setting, has a \SN\ in the range between 9 and 26\,per 1.3-\kms\ pixel in individual exposures. The \SN\ in the continuum near the \ion{Mg}{ii} doublet at $\approx$6030\,\AA\ in the red 580-nm setting is much higher, ranging between 28 and 62\,per pixel in individual exposures. The object is very bright so it was not necessary to observe it only during dark nights. Therefore, all exposures have moon illumination between $0$ and $100$\,per cent, with a mean of $43$\,per cent.         

  \begin{figure*}
  \centering
      \includegraphics[width=0.99\textwidth,natwidth=610,natheight=642]{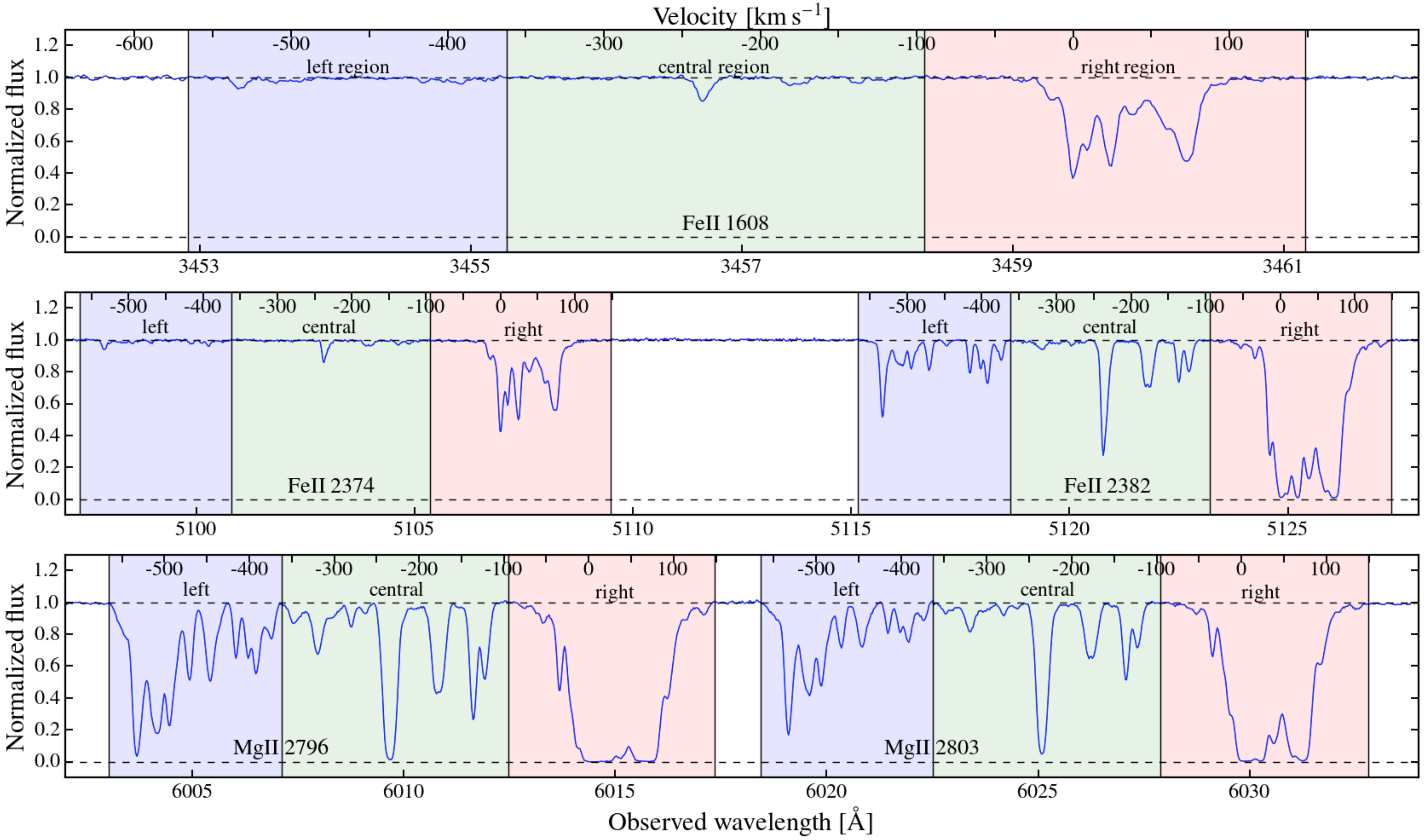}
  \caption{Example transitions from the combined UVES spectrum of HE 0515$-$4414. The upper panel shows the \ion{Fe}{ii}\,1608 transition, which represents the blue part of the spectrum, with ${\rm S/N}\sim 140$\,per 1.3-\kms\ pixel. The middle and lower panels show the \ion{Fe}{ii}\,2374 and 2382 transition pair and the \ion{Mg}{ii} doublet, respectively, with ${\rm S/N}\sim 240$\,per pixel; these represent the red part of the spectrum. Blue, green and red colours in each panel represent the three fitting regions in our analysis, referred to as the left, central and right region. The combined spectrum here includes all UVES exposures except those with 2$\times$1 binning and is available in \citet{srdan_kotus_2016_51715}.}\label{spectrum}
\end{figure*}
 
For data reduction we used the ESO UVES Common Pipeline Library (CPL 4.7.8). Initially it bias-corrects and flat-fields the quasar exposures. The quasar flux is then extracted with an optimal extraction method over the several pixels in the cross dispersion direction where the source flux is distributed. The wavelength calibration, which is a very important step for our measurements, was performed using an attached ThAr lamp exposure, taken immediately after each quasar exposure, and the air wavelengths and calibration procedure described by \citet{2007MNRAS.378..221M}. Instead of using the spectra which are automatically redispersed onto a linear wavelength scale by the CPL code, we used only un-redispersed flux and corresponding error arrays for individual echelle orders in the rest of reduction procedure.

After the wavelength calibration, the air wavelength scale of individual echelle orders, in all quasar exposures, was corrected to vacuum using the (inverse) Edlen (1966) formula. It was then converted to the Solar System heliocentric reference frame using the date and time of the mid-point of the exposure integration, using a custom code, {\sc uves\_popler} \citep{michael_murphy_2016_44765}. This code was also used to redisperse the flux from individual exposures onto a common log-linear wavelength scale with dispersion 1.3\,\kms\,$\text{pixel}^{-1}$ for 1$\times$1 binning and $2.5$\,\kms\,$\text{pixel}^{-1}$ for 2$\times$1-binned exposures. The rebinned flux arrays from all exposures were scaled to match that of overlapping orders and then combined with inverse-variance weighting and outlier rejection. 

We also used {\sc uves\_popler} to automatically fit a continuum to the spectrum using low-order polynomial fits to overlaping $2000$\,\kms\ wide sections. This continuum was generally acceptable, though some local adjustments were made using customised low-order polynomial fits in the vicinity of our transitions of interest. {\sc uves\_popler} automatically rejects pixels at the edges of echelle orders which have uncertainties in flux above some threshold, but we have rejected all pixels at the edges of echelle orders if they overlapped with our transitions of interest. We have also manually rejected some pixels from individual exposures around cosmic rays and other obvious artifacts. 

The final reduced spectrum, with all exposures combined except those with 2$\times$1 binning, is publicly available in \citet{srdan_kotus_2016_51715}. It covers all wavelengths from 3051 to 10430\,\AA\ except for a small inter-chip gap at 8537--8664\,\AA. Its resolving power is higher at redder wavelengths -- $\sim$63500 at $\la$4500\,\AA\ compared with $\sim$75000 at $\ga$5000\,\AA\ -- because a smaller slit width was typically used for the UVES red arm, though the many exposures contributing to the final spectrum had a range of resolving powers at all wavelengths, ranging between 53500--70000 at $\la$4500\,\AA\ and 62000--93500 at $\ga$5000\,\AA. We discuss the resolving power in more detail in \Sref{res_pow}. Example sections of the spectrum are shown in \Fref{spectrum} which cover 5 of the transitions used in our MM analysis. The continuum around the bluest transition, \ion{Fe}{ii}\,1608 shown in the top panel, falling near 3460\,\AA\ in the 346-nm setting, has a \SN\ of $\approx$140\,per 1.3-\kms\ pixel. The strongest transitions of \ion{Fe}{ii} and \ion{Mg}{i}/{\sc ii}, which dominate our analysis, fall in the wavelength range 5000--6400\,\AA\ (middle and bottom panels), are covered by many exposures in the 580-nm setting and all have a \SN\ of $\approx$240\,per pixel. The \SN\ peaks at $\approx$250\,per pixel around the \ion{Mg}{ii} doublet falling around 6015\,\AA. To our knowledge, these represent the highest \SN\ values for a quasar absorption system in an echelle spectrum at $z>1$.

Our spectra were observed over ten years, in five separate projects, with different charge-coupled device (CCD) on-chip binning and a variety of slit widths. This variety determines a range of different nominal resolving powers. This, combined with the very high $\SN$ of the spectra, means that modelling the absorption profiles accurately enough to measure $\varal$ is only possible if the combined spectrum is separated into five `sub-spectra' -- i.e.\ combined sub-sets of exposures taken in different `epochs'  with different resolving power and on-chip binning. These sub-spectra are summarised in \Tref{obs1}. Observations from project 60.A-9022(A) are separated into two sub-spectra, 1999a and 1999b, because exposures in 1999a have 2$\times$1 binning, while those in 1999b are not binned. Observations from projects 072.A-0100(A) and 079.A-0404(A) are combined together into sub-spectrum 0307 because they used the same slit width. Other observations were combined according to the project; we refer to these as sub-spectra taken in epochs 2000 and 0809. We do not notice any obvious deviations in the absorption profile shapes between the different epochs. However, this possibility could be investigated further with the published sub-spectra.

\subsection{HARPS/ESO-3.6\,m} \label{HARPS_ORC}
HARPS \citep{2003Msngr.114...20M} is a fiber-fed spectrograph with resolving power of $R\approx112000$. It is contained in an enclosure in which very stable conditions are maintained, such as very low and stable pressure and constant temperature. It is calibrated with ThAr lines and uses optical fibres, instead of a slit, to introduce light into the spectrograph. However, some of the systematic effects seen in the UVES and HIRES spectrographs also seem to be present in the HARPS wavelength scale. These systematic effects were first identified in the frequency comb study of \citet{2010MNRAS.405L..16W} as short-range distortions that are repeated from echelle order to echelle order. \citet{2013A&A...560A..61M} measured the amplitude of these `intra-order' distortions in each echelle order to be $\pm$40\,\ms. \citet{2013A&A...560A..61M} did not identify any significant long-range distortions, as were found subsequently in UVES and HIRES. While, \citet{2015MNRAS.447..446W} identified small $\sim$45\,\ms\,per 1000\,\AA\ long-range distortions in their analysis of HARPS solar twin spectra, they are most likely caused by systematic errors in the solar FTS spectrum used as the reference spectrum in that analysis. 

The HARPS spectrum used in this work consists of 18 exposures, with exposure times between 1\,h and 1.75\,h taken during six nights in 2003 and 2009 in ESO Projects 60.A-9036(A) and 072.A-0244(A), with a total exposure time of 93000\,s. A ThAr calibration exposure, taken before each night, was used to derive a wavelength solution for all quasar exposures on that night. The HARPS wavelength scale is stable to within just 15\,cm\,s$^{-1}$ over several hour time-scales \citep{2010MNRAS.405L..16W} so this approach does not limit the calibration uncertainty. The observations used one fibre on the quasar while the other was on a nearby sky position to allow sky subtraction. 

The quasar, sky and ThAr flux, together with an estimate of the blaze correction from flat-field exposures, were all automatically extracted by the standard HARPS data reduction software. This software also derived the wavelength calibration solution from the extracted ThAr flux. These products were then combined using {\sc uves\_popler} in the same way as the UVES exposures. However, the sky-subtraction, construction of a flux error spectrum, and blaze correction are not performed by the HARPS reduction software, so these steps were performed within {\sc uves\_popler}. The sky flux was redispersed onto the same common wavelength grid and subtracted from the quasar flux in the same echelle order before the (sky-subtracted, blaze-corrected) flux from all orders and exposures was combined. The error spectrum for each exposure was estimated, assuming Gaussian statistics, from the combination of the quasar CCD electron counts and an estimate of the read noise in each extracted pixel.

The final reduced HARPS spectrum that we use in our analysis, which we make publicly available in \citet{srdan_kotus_2016_51715}, covers the wavelength range between 3791 and 6905\,\AA\ with a small gap between 5260 and 5338\,\AA\ due to the physical gap between the HARPS blue and red CCD chips. The \SN\ around 5050\,\AA\ is $\approx$29\,per 0.85-\kms\ pixel but at the position of the \ion{Mg}{ii} doublet it peaks at $\approx$33\,per pixel.

\subsection{bHROS/Gemini} \label{bHROS_ORC}

bHROS \citep{2008psa..conf..297M} was bench-mounted in the gravity-invariant pier of the Gemini South Telescope and fed by a 0\farcs9\,$\times$\,0\farcs9 optical fibre that, via an image slicer, projected a 0\farcs14-wide pseudo-slit into the spectrograph to achieve a resolving power of $R\approx140000$. A total of 37\,$\times$\,3600-s exposures of HE0515$-$4414 were obtained during bHROS ``science verification" in November and December 2006 and January 2007 during variable conditions. Each quasar exposure was followed immediately by a ThAr lamp exposure. Unfortunately, one of the two CCD chips failed before the observations and the efficiency of bHROS appeared well below specifications during them. The former meant that observations in 2 separate wavelength settings were required to cover most strong transitions in the $\zab=1.1508$ absorber. The spatial profile of light from the image slider was spread over a large number of pixels ($\approx$50), so the contribution of CCD read noise was significant and further reduced the \SN\ of the spectra considerably. ThAr lamp observations also revealed instabilities that caused $\sim$0.5\,\kms\ shifts in the spectrum over several hours. These factors dramatically reduced the \SN\ of the spectra and rendered them useless for directly measuring $\varal$. However, the very high resolution may still assist in revealing additional velocity structure, and we explore this possibility in \Sref{sys_v_str}.

No dedicated data reduction pipeline is available for bHROS spectra, so we used our own custom codes, based loosely on the {\sc reduce} suite of routines \citep{2002A&A...385.1095P}. The two-dimensional shifts between quasar exposures was determined from their corresponding ThAr exposures and the heliocentric velocity at the mid-point of the quasar exposure integration. These shifts were used to combined the low-\SN, individual (dark current, bias and flat-field corrected) quasar exposures into a single, higher-\SN\ one upon which the flux extraction procedure was performed. The flux in each echelle order containing a transition of interest was extracted with a custom optimal extraction routine which accounted for bad pixels and cosmic rays. Error arrays were estimated in the same process. No sky-subtraction was possible because only a single fibre was available. However, the observations were conducted with typically $\sim$25--50\,per cent moon illumination, so the sky background flux is expected to be negligible.

The final, reduced bHROS spectrum is publicly available in \citet{srdan_kotus_2016_51715} for the transitions of \tran{Mg}{i}{2852}, \tran{Mg}{ii}{2796}, 2803 and \tran{Fe}{ii}{2383}, 2586 \& 2600 that could be covered in two separate wavelength settings. The spectra around these transitions were extracted to a linear wavelength grid with 0.012\,\AA\ dispersion and the final \SN\ for, e.g., the \tran{Fe}{ii}{2382} at 5123\,\AA\ is $\approx$24 per 0.70-\kms\ pixel.

\section{Analysis}

\subsection{Identification of absorption systems and transitions}\label{abs_sys}
To measure $\varal$ we may use all available transitions for which the $q$ coefficients are known and for which the rest wavelengths are measured with high enough precision. Such transitions were reviewed in \citet[][]{2014MNRAS.438..388M} and those used in our analysis are summarised in \Tref{resolving}. We only constrain $\varal$ in the absorption system at $z=1.1508$. However, we need to identify absorption systems at other redshifts that overlap with those of the $z=1.1508$ system to understand the effect of their transitions, if any.

For absorption system identification we used a custom-made code which searches for absorption from different metal transitions of different species in redshift space. We have identified 8 other absorption systems towards QSO HE 0515$-$4414 at redshifts 0.2223, 0.2818, 0.4291, 0.9406, 1.3849, 1.5145, 1.6737 and 1.6971. 

In the absorption system of interest the strongest transitions show very wide ($\sim$750\,\kms) and very complex velocity structure (e.g. bottom panel of \Fref{spectrum}) providing a reasonable probability of a blend with transitions from these other redshifts. We have identified 3 such blends. \ion{Al}{ii}\,1670 at $z=1.1508$ is blended with \ion{Mg}{ii}\,2803 at $z=0.2818$, and because of that we do not fit \ion{Al}{ii}\,1670 redwards of $-140$\,\kms\ (see \Sref{fitt}). \ion{Al}{iii}\,1862 at $z=1.1508$ is blended with \ion{Mg}{ii}\,2803 at $z=0.4291$, so we do not fit \ion{Al}{iii}\,1862 redwards of $-100$\,\kms. \ion{Fe}{ii}\,2344 at $z=1.1508$ is blended with \ion{Ca}{ii}\,3934 at $z=0.2818$, hence the former is not fitted above $70$\,\kms. We also do not fit parts of other weak transitions where we do not formally detect any absorption (e.g. \ion{Fe}{ii}\,1611).

\begin{table}
\caption{Spectral resolution for different sub-spectra and transitions used in the $\varal$ analysis. Specific FWHMs for each transition (its wavelength and echelle order position) are estimated from the FWHMs of ThAr lines observed in the attached exposures.}\label{resolving}
\centering
  \begin{tabular}{@{}c c c c c}   
  \hline
 Ion Transition  & \multicolumn{4}{c}{FWHM [\kms]}                                   \\
                 & 1999       & 2000          & 0307          & 0809           \\
 \hline
  \ion{Mg}{i}\,2026    &  5.308       &  4.363        &  no spectra     &  no spectra       \\
  \ion{Mg}{i}\,2852    &  4.199       &  4.670        &  4.736          &  3.768            \\
  \ion{Mg}{ii}\,2796    &  4.262       &  4.714        &  4.818          &  3.876            \\
  \ion{Mg}{ii}\,2803    &  4.234       &  4.696        &  4.780          &  3.824            \\
  \ion{Al}{ii}\,1670    &  5.296       &  4.660        &  4.361          &  4.340            \\
  \ion{Al}{iii}\,1854    &  5.395       &  4.349        &  no spectra     &  no spectra       \\
  \ion{Al}{iii}\,1862    &  5.597       &  4.394        &  no spectra     &  no spectra       \\
  \ion{Si}{ii}\,1808    &  5.423       &  4.356        &  no spectra     &  no spectra       \\
  \ion{Cr}{ii}\,2056    &  5.223       &  4.385        &  no spectra     &  no spectra       \\ 
  \ion{Cr}{ii}\,2062    &  5.433       &  4.389        &  no spectra     &  no spectra       \\
  \ion{Cr}{ii}\,2066    &  5.261       &  4.354        &  no spectra     &  no spectra       \\
  \ion{Mn}{ii}\,2576    &  3.933       &  4.366        &  4.302          &  3.309            \\
  \ion{Mn}{ii}\,2594    &  3.850       &  4.338        &  4.227          &  3.212            \\
  \ion{Mn}{ii}\,2606    &  3.900       &  4.340        &  4.269          &  3.264            \\
  \ion{Fe}{ii}\,1608    &  5.310       &  4.667        &  4.363          &  4.396            \\  
  \ion{Fe}{ii}\,1611    &  5.263       &  4.728        &  4.361          &  4.385            \\
  \ion{Fe}{ii}\,2260    &  3.848       &  4.359        &  4.295          &  3.294            \\
  \ion{Fe}{ii}\,2344    &  3.841       &  4.361        &  4.300          &  3.311            \\
  \ion{Fe}{ii}\,2374    &  3.829       &  4.356        &  4.290          &  3.283            \\
  \ion{Fe}{ii}\,2382    &  3.841       &  4.366        &  4.305          &  3.316            \\
  \ion{Fe}{ii}\,2586    &  3.881       &  4.323        &  4.246          &  3.233            \\
  \ion{Fe}{ii}\,2600    &  3.928       &  4.366        &  4.302          &  3.311            \\
  \ion{Ni}{ii}\,1709    &  5.374       &  4.696        &  4.359          &  4.302            \\
  \ion{Ni}{ii}\,1741    &  5.595       &  4.780        &  4.356          &  4.269            \\
  \ion{Ni}{ii}\,1751    &  5.331       &  4.519        &  4.356          &  4.269            \\
  \ion{Zn}{ii}\,2026    &  5.327       &  4.368        &  no spectra     &  no spectra       \\
  \ion{Zn}{ii}\,2062    &  5.411       &  4.385        &  no spectra     &  no spectra       \\
   \hline
   \end{tabular}
  \end{table}

\subsection{Modelling the resolving power} \label{res_pow}
Even small inaccuracies in the estimated resolving power can yield statistically significant differences between the profile model and spectral data, especially when fitting a large number of velocity components in very high $\SN$ spectra. Therefore, it is very important to accurately estimate the resolving power. Studies of even the best spectra \citep[e.g.][]{2014MNRAS.445..128E} have previously been able to simply assume nominal resolving power values. However, our spectrum has such high $\SN$ that a more sophisticated approach is needed. Further, even the ``quality control" information available from ESO (plots of resolving power versus slit width) was not accurate enough for our purposes, because the resolving power can vary by up to 20 per cent depending on the position along the echelle order, the seeing, and/or the slit width. 

To obtain a more accurate first-guess resolution, we estimated the resolving power for each transition in each sub-spectrum from the FWHM of ThAr lines used in our calibration. We performed a linear fit to FWHM versus wavelength and FWHM versus position along the echelle orders of the different chips in each sub-spectrum. From these fits we calculated the FWHM of ThAr lines at the specific wavelengths in each sub-spectrum for our transitions of interest. A similar, but to some extent simplified approach was used in \citet{2013A&A...555A..68M}.

Quasar and ThAr light have slightly different light paths through the optics of the spectrograph. Quasar light is blurred by the atmosphere, forming a seeing disc at the slit entrance. We model this in the spectral direction as a Gaussian function, with a FWHM equal to the seeing. The seeing disc is truncated by the edges of the slit jaws, which we model in the spectral direction as a hat function. We therefore represent the quasar light after the slit as the product of the Gaussian (seeing) and hat (slit) functions. On the other hand, ThAr light fully illuminates the slit, so it can be modelled just with the hat function. Both signals subsequently pass through similar optics, so to represent their 1-dimensional point-spread functions at the CCD, we convolve them with the same instrumental profile, represented by a Gaussian with a 0.13 arcsec width, which is smaller than the smallest slit width of UVES. 

To understand the difference between the quasar and ThAr resolving powers, we calculate the ratio of the measured FWHMs of these final two 1-dimensional profiles. We found this ratio of ThAr FWHM and QSO absorption line FWHM to be between 0.8 and 1. However, in this modelling process we did not account for two-dimensional effects or additional blurring caused by imperfect tracking of the quasar. The latter will be largest for very small values of the seeing-to-slit ratio, because in that situation the observer cannot see any part of the seeing disc reflected from the slit jaws to track the quasar’s position. Therefore, we have imposed a lower limit of 0.9 for the ratio of ThAr-to-quasar resolving powers. The final resolving powers for the different sub-spectra and transitions are shown in \Tref{resolving}.

After refining the resolving power for each transition in the above way, the $\chisq$ of our fits to their absorption profiles improved even before rerunning the $\chisq$ minimization procedure again. This justifies the procedure and should be taken into account in future measurements with high $\SN$ and should possibly involve taking into account two dimensional effects, which will improve the fits further.

\subsection{Correcting the UVES wavelength scale with HARPS spectra}\label{DC_method}
To establish the correct wavelength scale and to measure $\varal$ accurately, long-range distortionsin the wavelength scales of the UVES spectra need to be accounted for. \citet{2014MNRAS.445..128E} did this by supercalibrating asteroid spectra (reflected sunlight), taken with the same instrument as their quasar spectra, with Fourier transform spectra (FTS) of the Sun. \citet{2015MNRAS.447..446W} demonstrated that similar supercalibration can be accomplished using spectra of solar twin stars. However, the UVES spectra of HE 0515$-$4414 were observed over many widely-separated epochs, mostly in service mode, without appropriate solar twin stars or asteroid exposures within the same nights. Therefore, we must use a different approach and, fortunately, due to the quasar's brightness, the HARPS spectra are of sufficient quality to provide a similar supercalibration. 

The solar twin supercalibration of HARPS by \citet{2015MNRAS.447..446W} revealed a $\sim$45\,\ms\,per 1000\,\AA\ long-range distortion. This is consistent with the first asteroid supercalibration results for HARPS by \citet{2011A&A...525A..74M}, though that study employed fewer lines and the long-range distortion was less clear. However, in the frequency comb calibration of HARPS by \citet{2013A&A...560A..61M}, no long-range distortion was apparent. It is therefore likely that the long-range distortion seen in \citet{2015MNRAS.447..446W} results from a small distortion in the solar FTS spectra used and not HARPS itself. Therefore, in this work we regard the HARPS spectra as having no long-range distortion.

Our approach is to directly compare the HARPS spectrum of the same quasar to its UVES counterpart to transfer its much more accurate wavelength scale. This will correct the long-range distortions in the UVES spectrum's wavelength scale and allow an accurate $\varal$ measurement which utilizes the high S/N of the UVES spectrum. In other words, we use the Direct Comparison (DC) method \citep{2013ApJ...778..173E} to transfer the HARPS wavelength scale, which is more accurate, to the UVES spectra.

The DC method compares corresponding small sections (``chunks'') of two spectra and robustly measures the velocity shift between them using all available absorption features which contribute information beyond some specified significance threshold (typically 5--10$\sigma$). It is used to find the weighted mean shift between the sections and any change in this shift over long wavelength ranges. A velocity shift may be caused by the quasar being positioned differently across the slit during different exposures and this shift should not vary substantially with wavelength. We refer to these as ``slit-shifts'' below. However, it is not known what specifically causes the long-range distortions identified in \citet{2013MNRAS.435..861R}  and explored by \citet{2015MNRAS.447..446W} in detail, but, in general terms, they are likely caused by the differences in the light paths of the quasar and the ThAr lamp light through the spectrograph. In the following subsections we use the DC method to find velocity shifts and slopes (stretch or compression of the wavelength scale) between different UVES exposures in each sub-spectrum (\Sref{DC_ind}), between the combined UVES sub-spectra from different epochs (\Sref{DC_epoch}), and between the combined UVES sub-spectrum from each epoch and the HARPS spectrum of the same quasar (\Sref{DC_HARPS}).

\subsubsection{Velocity shifts between UVES exposures within each sub-spectrum} \label{DC_ind}
To measure and correct for slit-shifts between exposures within a given sub-spectrum, we used the DC method to compare each of them with the combined sub-spectrum from that epoch. The DC method was applied using 200-\kms\ chunks and a feature-selection threshold of 7$\sigma$. These are similar to the parameters used by \citet{2013ApJ...778..173E} for non-Lyman-$\alpha$ forest regions of spectra, with a slightly higher selection threshold because of the much higher \SN\ of our spectra. In principle, the individual exposures within each sub-spectrum may have long-range distortions between them which should first be measured and corrected before the slit-shifts. However, from the DC method comparison, we found that these distortions had slopes that were all consistent with zero, albeit with large statistical uncertainties (typically $\sim$190\,\ms\,per 1000\,\AA) because of the relatively low \SN\ per exposure ($\la$40\,per pixel). Therefore, we did not correct for long-range distortions and proceeded to measure and correct the slit-shifts only. 

We found shifts typically in the range between $-100$ and 100\,\ms, with extreme values of 300 \ms, which is consistent with both cross-correlation measurements of \citet{2013MNRAS.435..861R} and DC method measurements of \citet{2014MNRAS.445..128E}. It is very important to correct for slit-shifts because they can influence how the spectra are combined within {\sc uves\_popler}. For example, if the \SN\ varies differently as a function of wavelength in two different exposures, the combined spectrum will have a wavelength-dependent relative contribution from those exposures and, thus, a wavelength-dependent residual slit-shift. Transitions at different wavelengths will therefore have a spurious velocity shift between them. Therefore, the slit-shift for each exposure was corrected within {\sc uves\_popler}, by applying the opposite of the shifts to all exposures in every sub-spectrum. After re-combination, there were no shifts left among different exposures in each sub-spectrum. This was checked with a repeated DC method analysis.

\subsubsection{Velocity shifts between UVES sub-spectra} \label{DC_epoch}
At this stage, individual exposures that are combined to create the sub-spectra had been corrected for the slit-shifts. However, different sub-spectra could still have shifts between them. This kind of shift is even more important to correct for, due to the different wavelength coverage of each sub-spectrum. For example, some sub-spectra do not contribute to some transitions, while other sub-spectra contribute to all transitions. If there is a shift between two such sub-spectra, it will directly influence the velocity shift between these two sets of transitions, leading directly to a spurious $\varal$ measurement.

The higher S/N of the sub-spectra (compared to their constituent exposures) also allows us to estimate relative wavelength distortions between them. These can be checked against the absolute (and more precise) long-range wavelength distortions measured in \Sref{DC_HARPS}. We have chosen to compare all the sub-spectra with the 2000 sub-spectrum, because it consists of exposures in all the different settings that we use in our analysis and therefore has the largest wavelength coverage. We used the same DC method parameters as in the previous subsection. 

We found slit-shifts between sub-spectra in the range between $-200$ and 200\,\ms, consistent with shifts found in previous works of \citet{2013MNRAS.435..861R} and \citet{2014MNRAS.445..128E} and our shifts found in \Sref{DC_ind} . We corrected for these relative shifts and recombined the sub-spectra using {\sc uves\_popler} by adding the additional correction to the slit-shift for each exposure from \Sref{DC_ind}. Upon re-combination, no shifts remained between any of the exposures, which was again checked with a repeated DC method analysis.

The relative long-range distortions of all the sub-spectra, except one, have slopes that are statistically consistent with zero, with uncertainties of typically $\sim$30 and in the range $\sim$170--560\,\ms\,per 1000\,\AA, for the red and blue settings, respectively, and an average of $\approx$200\,\ms\,per 1000\,\AA. This average uncertainty is typical of the magnitude of the distortions found in UVES spectra \citep{2015MNRAS.447..446W}. Assuming a simple MM analysis involving only the \ion{Mg}{ii}\,2796 and \ion{Fe}{ii}\,2382 transitions, this would correspond to an uncertainty in $\varal$ of $\approx$10\,ppm. This is still an order of magnitude larger than the final systematic error budget of $\la$1\,ppm which we measure in this work (see \Sref{results}). In a single exceptional case, the 0307 red 580-nm setting, the long-range distortion relative to the 2000 sub-spectrum had a slope of 119$\pm$30\,\ms\,per 1000\,\AA. If converted to $\varal$ in the same way as above, it would represent a $\varal$ of $\sim$6\,ppm. That is, left uncorrected, such a distortion could contribute a significant systematic error. We return to this case in \Sref{DC_HARPS} below.

\subsubsection{Correction of each UVES sub-spectrum to the HARPS wavelength scale} \label{DC_HARPS}
Finally, we directly compared the combined UVES sub-spectra with the HARPS spectrum. The HARPS spectrum has the disadvantage of limited wavelength coverage, 3780--6915\,\AA, compared to 3040--10430\,\AA\ for the UVES spectrum, so only the limited, overlapping wavelength range of the two spectra can be compared. This corresponds to the wavelengths covered by the UVES 580-nm setting. Therefore, we cannot determine the size of any velocity shift between the red and blue arms of UVES from our spectra and/or whether the long-range distortions in the red and blue arms are the same. In the analysis of \citet{2015MNRAS.447..446W}, the slopes of the distortions in UVES's red and blue arms were indeed found to be very similar in most cases, so we make this assumption below.  We assume that the distortion slopes from the UVES 580-nm setting applies to all other UVES settings in the 1999, 0307 and 0809 sub-spectra. We test this assumption in \Sref{without_blue}.

The slopes of the distortions we find between the HARPS spectrum and the UVES sub-spectra are provided in \Tref{slopes} and fall between 110 and 220\,\ms\,per 1000\,\AA.  The larger slopes correspond to $\varal\approx10$\,ppm if we consider the same two transitions used in \Sref{DC_epoch} for the conversion, so it is clearly important to correct for these distortions. This also indicates the size of the systematic error expected in previous analyses of the 1999 and 2000 spectra. The distortions in \Tref{slopes} are comparable to the relative distortions found in the DC method comparison of the UVES sub-spectra from different epochs in \Sref{DC_epoch}. They are typical of the long-range distortions found in UVES spectra for the epochs probed \citep{2015MNRAS.447..446W}.

Another relevant disadvantage of the HARPS spectrum is its much lower S/N compared to the UVES spectrum (e.g. $\approx$35\,per 0.85\,\kms\ pixel at 6000\,\AA), and this limits the precision of the slope determined from the DC method analysis. However, the statistical uncertainties in the distortion slopes reported in \Tref{slopes} are relatively small, between 35 and 53\,\ms\,per 1000\,\AA, meaning that the distortions themselves are detected with 2.2--6.0$\sigma$ significance.

While the 1999, 2000 and 0809 sub-spectra all have similar distortion slopes, the slopes found for the 0307 and 2000 sub-spectra differ by $\approx$109\,\ms\,per 1000\,\AA. This is consistent with the relative difference found by directly comparing the UVES 0307 and 2000 sub-spectra in Section 3.3.2. This demonstrates the robustness of the DC method and HARPS supercalibration approach, and increases confidence in the final wavelength scale of our UVES spectrum.  We have therefore corrected for the distortions listed in Table 3 by applying an opposing long-range distortion slope to the each corresponding exposure in {\sc uves\_popler}. We use the error bars associated with these slopes to estimate the remaining systematic uncertainty budget in \Sref{long-range_sys}.

\begin{table}
\caption{Long-range distortion slopes identified in the DC method comparison of different UVES sub-spectra in the 580-nm setting with the HARPS spectrum. It is important to correct for the long-range wavelength distortions with this magnitude because they can lead to spurious $\varal$ measurements of up to $\sim$10\,ppm.}\label{slopes}
\centering
  \begin{tabular}{@{}c c}   
  \hline
 Sub-spectrum        &   Slope [$\ms\,\angstrom^{-1}$]	      \\
 \hline
1999 &  $0.209\pm   0.035$   \\
2000 &  $0.221\pm   0.053$   \\ 
0307 &  $0.112\pm   0.051$   \\
0809 &  $0.203\pm   0.041$   \\
  \hline 
  \end{tabular}
  \end{table} 

After correcting for the long-range wavelength distortions, the DC method was run again to determine any newly-introduced velocity shifts between the corrected UVES sub-spectra and the HARPS spectrum. These shifts will be introduced because the different UVES sub-spectra cover different wavelength ranges and the slopes are corrected around different `pivot' wavelengths. That is, they are artifacts of our procedure. They are found to be $<$185\,\ms\ corrections. Shifts were measured for the 437, 520 and 580-nm settings, but for the 346-nm setting we used the shifts from the 580-nm setting from the same sub-spectrum. This should be correct if the previous assumption of similar long-range wavelength slopes in the blue and red settings is valid, and we test this in \Sref{without_blue}. We also checked that the 437 and 580-nm settings have consistent shifts and we found them to be consistent. However, the uncertainties on the shifts in the 437-nm settings are much larger than for the 580-nm settings. Therefore, we cannot make a definitive conclusion that these shifts, as well as the long-range wavelength distortion slopes, are similar in the blue and red settings from this information alone. The final velocity shift corrections were added to the previously measured offsets in all exposures from the relevant sub-spectra and they were recombined with {\sc uves\_popler}. 

As a last step in this part of the analysis, we have confirmed that a DC method analysis returns negligible, insignificant shifts and distortion slopes between the UVES sub-spectra and the HARPS spectrum after all of these corrections.  

\subsection{Fitting procedure}\label{fitt}
Due to the very complex and broad velocity structure of the $\zab=1.1508$ absorber, we have separated the absorption profile into three regions, which we fit separately: `left',`central' and `right'. These velocity regions are defined such that there appears to be no absorption in the strongest transition (\ion{Mg}{ii}\,2796) at the region edges, and that they have approximately the same velocity width. The regions are indicated by different shading in \Fref{spectrum}, where left, central and right regions are between $-$565\,\kms\ and $-$360\,\kms, $-$360\,\kms\ and $-$100\,\kms, $-$95\,\kms\ and 150\,\kms, respectively, where $v=0$\,\kms\ is at $z=1.1507930$. If there is no absorption in some parts of a weak transition its regions are narrowed to avoid fitting many pixels of unabsorbed continuum unnecessarily.

To be able to measure $\varal$ we need to construct a model of the absorption profile, jointly for all fitted transitions, which comprises many individual velocity components. These are represented as Voigt profiles convolved with Gaussian instrumental profiles. We use {\sc vpfit} version 9.5 \citep{2014ascl.soft08015C} to minimize the $\chisq$ between this model and the spectra. This is a non-linear least-squares algorithm which varies the column densities ($N$), Doppler $b$-parameters and redshifts ($z$) of individual components. We allow column densities to vary freely, but we tie $b$-parameters and redshifts for each velocity component between different transitions. We assume that all velocity components in our fits are turbulently broadened, i.e.\  the $b$-parameters of corresponding components are the same for all species. However, previous work has demonstrated that $\varal$ is not sensitive to the broadening mechanism assumed, either purely thermal (where the $b$-parameter scales with the atomic mass) or a mix of turbulent and thermal \citep[e.g.][]{2003MNRAS.345..609M,2012MNRAS.422.3370K}. This is especially the case with a fitting approach that seeks to fit all statistically significant structure \citep{2014MNRAS.445..128E}, so we expect that systematic errors from the assumption of turbulent broadening are negligible. {\sc vpfit} can also vary the additional $\varal$ parameter, but initially we fix $\varal$ to zero until our preferred models of the absorption profiles are finalized.

Before fitting we blinded our spectra by introducing random small shifts between transitions in {\sc uves\_popler}. This blinding approach was the same as employed by \citet{2014MNRAS.445..128E}: random intra-order and long-range distortions were introduced into each exposure, with some common elements between exposures (to prevent the distortions averaging out over many exposures). The amplitudes and magnitudes of the distortions were such that no value of $\varal$ derived from the sub-spectra could be trusted, but not large enough to significantly affect the profile fitting approach below. The blinding was only removed once the preferred profile fits were finalized and $\varal$ was to be determined.

\begin{figure*}
  \centering
    \includegraphics[width=0.93\textwidth,natwidth=610,natheight=642]{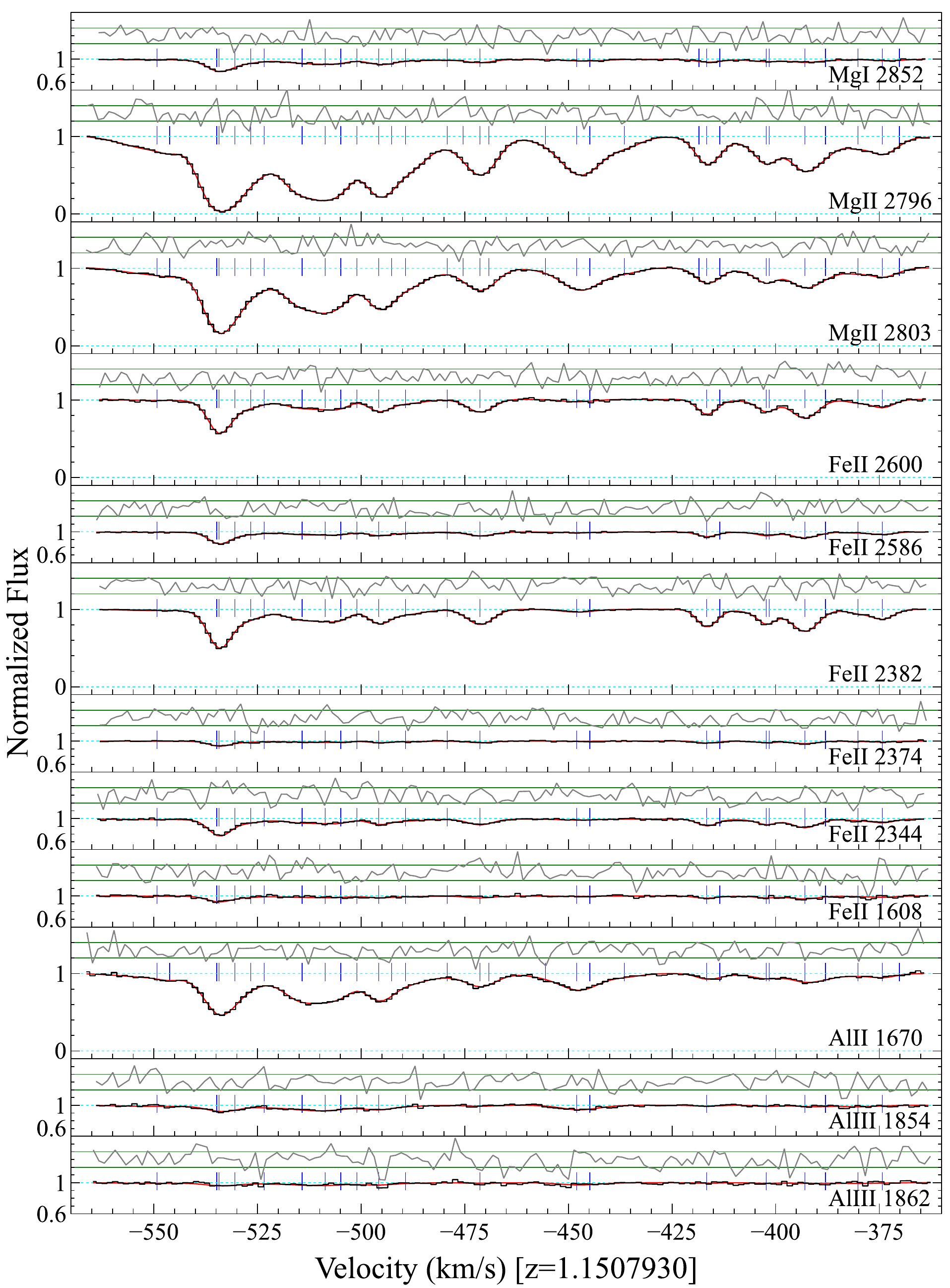}\vspace{-1em}
   \caption{Combined, continuum-normalized UVES 0809 sub-spectrum [ESO project 082.A-0078(A)] (black histogram) overlaid with our preferred model (red curve) in the left region. For transitions not covered by the 0809 sub-spectrum (see \Tref{resolving}), the 2000 sub-spectrum is plotted. Blue tickmarks represent centroids of individual velocity components. Dotted light blue lines below and above the spectra represent the zero level and continuum, respectively. The grey line represents the residuals between the spectrum and the model, with the $\pm1\sigma$ deviations marked by the dark green lines. The fitting region is narrowed down in weak transitions due to the lack of absorption, or they are not fitted at all.}\label{left_a}
\end{figure*}
\begin{figure*}
  \centering
    \includegraphics[width=0.99\textwidth,natwidth=610,natheight=642]{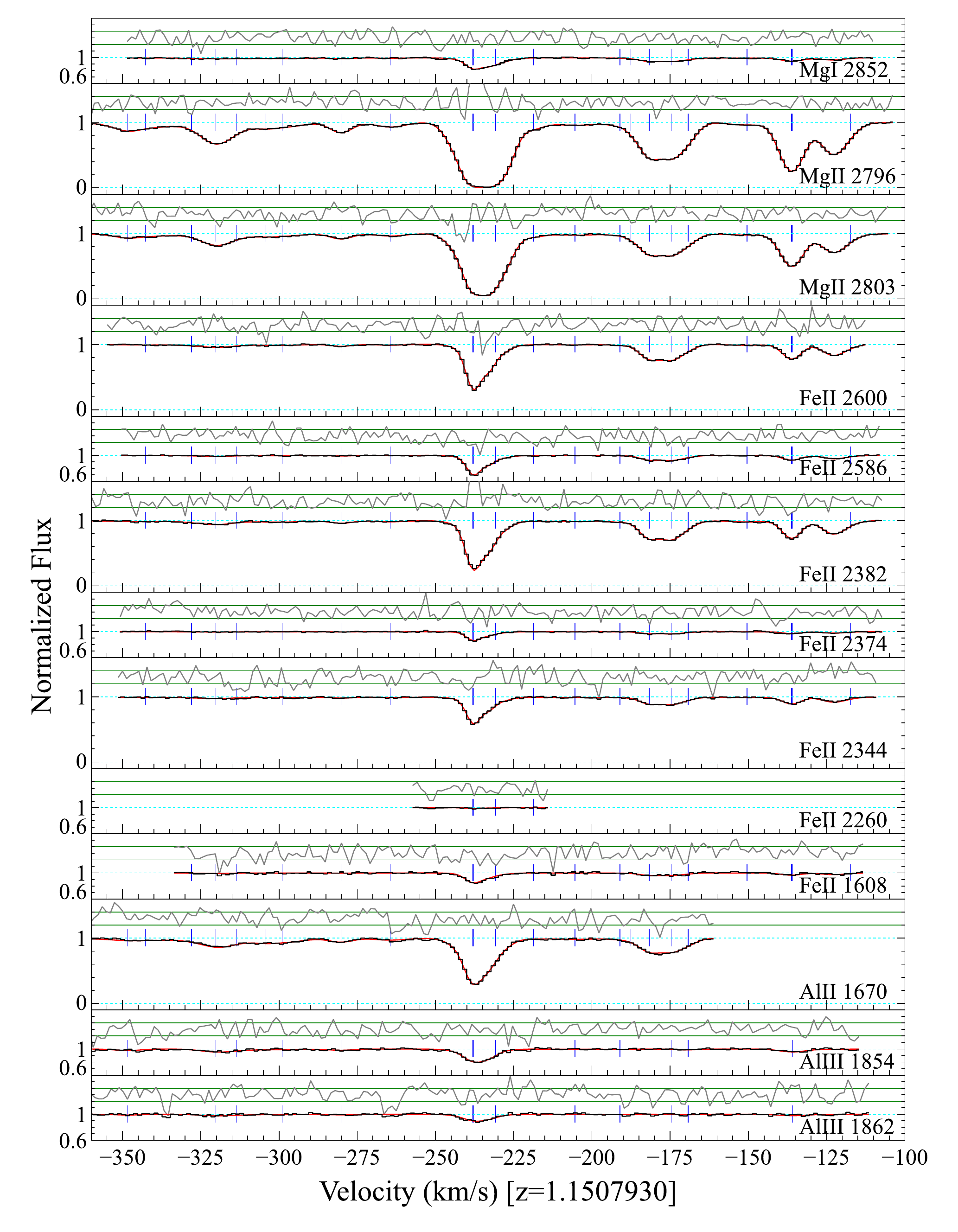}
  \caption{Same as \Fref{left_a} but for the central region. The fitting region for \ion{Al}{ii} is not modeled above $-$160\,\kms\ because of the blend discussed in \Sref{abs_sys}.}\label{cent_a}
\end{figure*}

\begin{figure*}
  \centering
    \includegraphics[width=0.99\textwidth,natwidth=610,natheight=642]{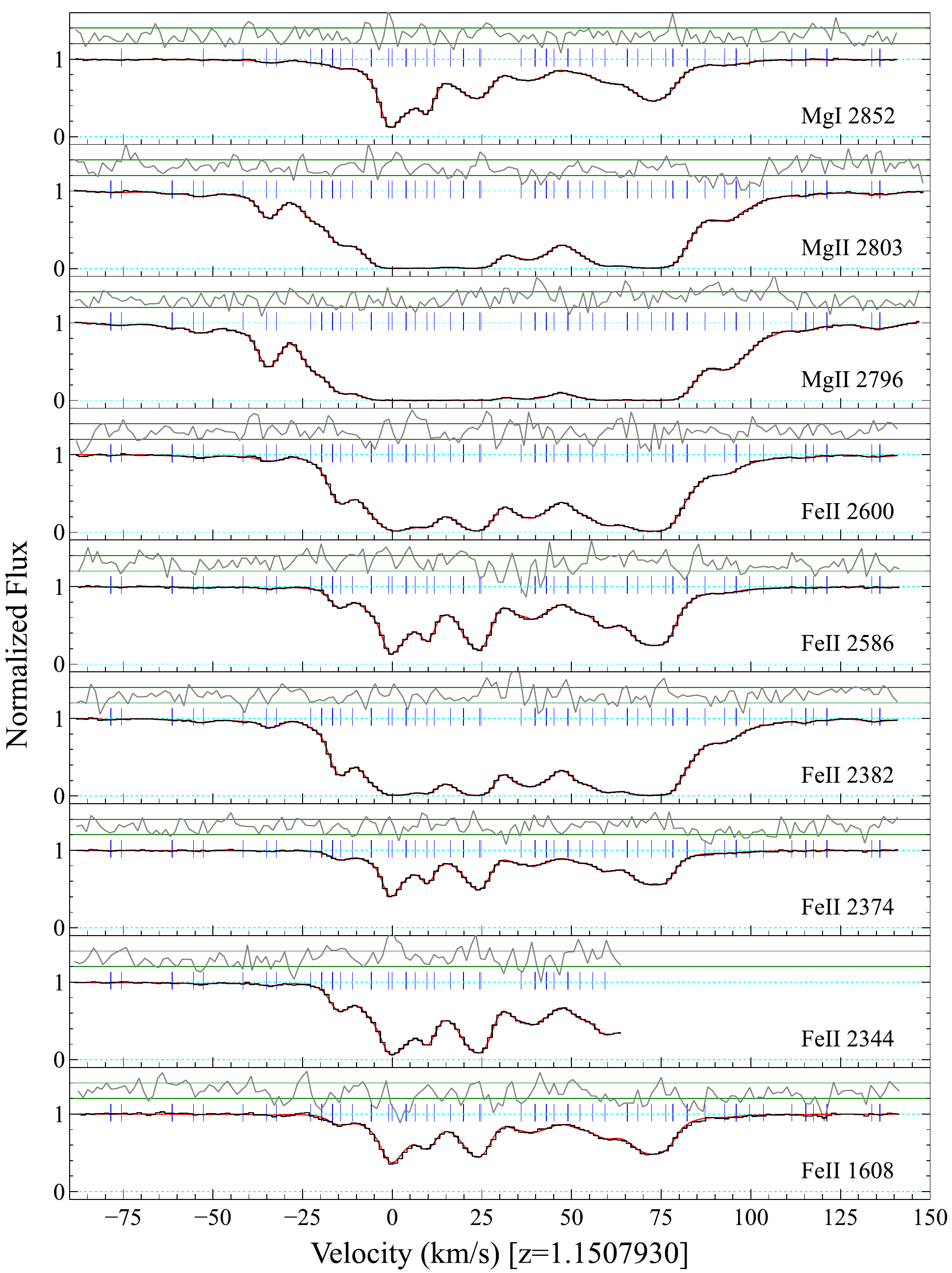}
  \caption{Same as \Fref{left_a} but for the strong transitions in the right region. The fitting region for \ion{Fe}{ii}\,2344 is not modeled above 65\,\kms\ because of the blend discussed in \Sref{abs_sys}.}\label{right_a}
\end{figure*}
\begin{figure*}
  \centering
    \includegraphics[width=0.99\textwidth,natwidth=610,natheight=642]{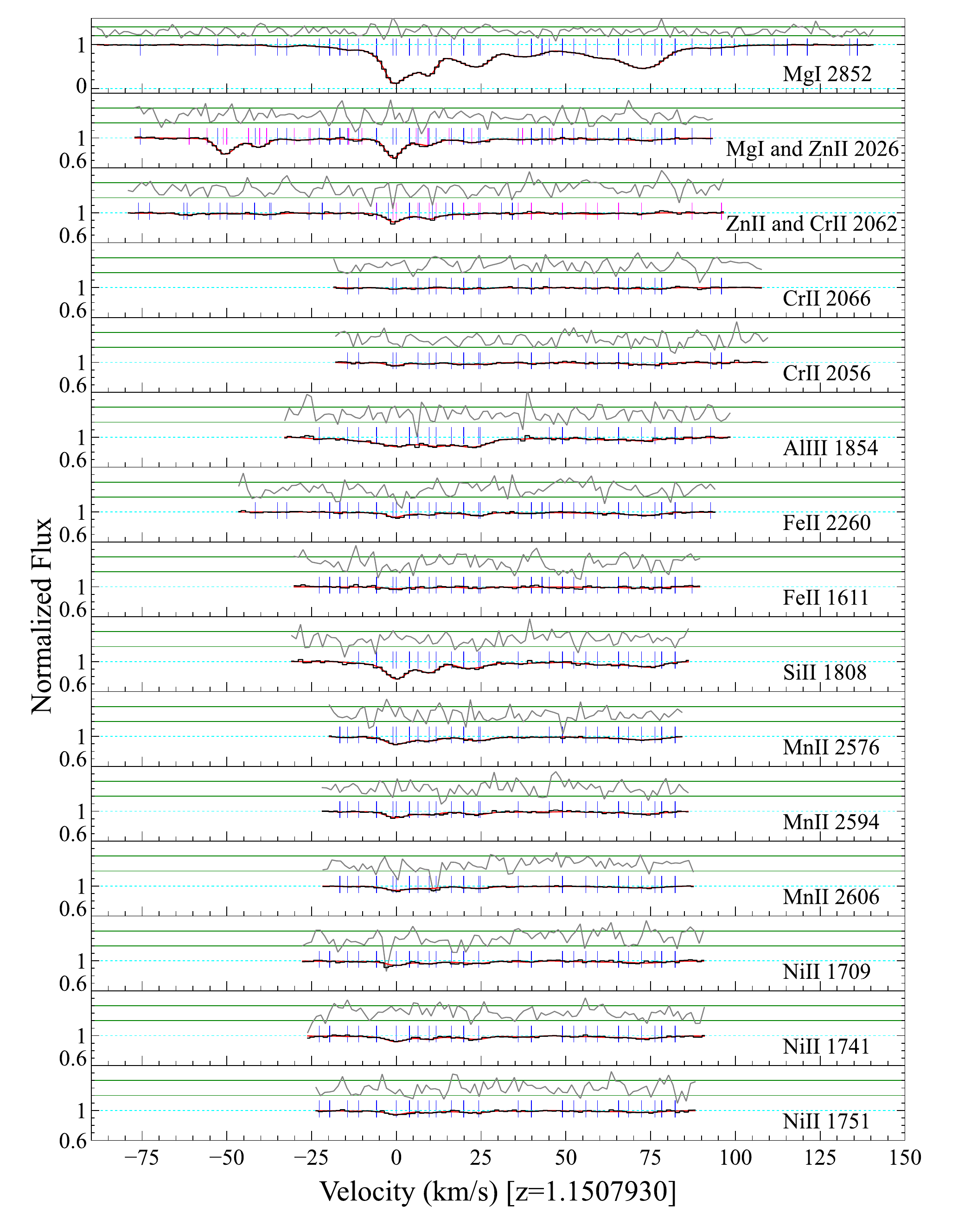}
  \caption{Same as \Fref{left_a} but for the weak transitions in the right region. Pink tickmarks in the second and third panel represent velocity components of \ion{Zn}{ii} transitions and blue tickmarks are components of \ion{Mg}{i}\,2026 and \ion{Cr}{ii}\,2062 transitions.}\label{right_b}
\end{figure*}

Our approach to fitting is similar to recent analyses of variation in $\alpha$ \citep[e.g.][]{2013A&A...555A..68M, 2014MNRAS.445..128E}. However, we had to divide the fitting process into several steps due to the high complexity of the absorption system. Firstly, we fitted just the \ion{Fe}{ii} transitions because they cover a variety of optical depths (because they have a variety of oscillator strengths). In an iterative process we trialed many models by reducing $\chisq$ using {\sc vpfit} and adding more components in places where the residuals were not satisfactory. Once a statistically satisfactory fit was established using just the \ion{Fe}{ii} transitions we incorporated the weaker transitions, one by one, by using the same velocity structure as in \ion{Fe}{ii} and minimizing $\chisq$ in {\sc vpfit}. In this process we also narrowed the velocity width of regions in parts of the weak transitions where we did not identify any absorption. In all such weak transitions (i.e.~those of \ion{Zn}{ii}, \ion{Cr}{ii} etc.), the weaker velocity components were not typically required to provide a statistically acceptable fit, so {\sc vpfit} removed them. We were careful to check each time such a component was removed that it corresponded to the weakest of any close grouping of \ion{Fe}{ii} components. Typically, many fewer components are required to fit these much weaker transitions, but our fit maintains the components corresponding to the strongest \ion{Fe}{ii} components. After inspecting the residuals again, we tried to incorporate additional components if possible. Finally, we incorporated the somewhat saturated \ion{Mg}{ii} transitions by copying the \ion{Fe}{ii} velocity structure, including some additional velocity components and minimizing $\chisq$, iteratively trialing more complex models until we achieved the statistically preferred model for $\varal$ measurements.   

The 5 different sub-spectra have somewhat different resolving powers. Therefore, rather than fitting the combined spectrum, the sub-spectra were fit separately but simultaneously with the same model. For each sub-spectrum the model is convolved with a Gaussian line-spread function with the appropriate FWHM shown in \Tref{resolving}.

Our preferred models of the absorption profiles are shown in Figs.~2--5 overlaid on the 0809 sub-spectrum -- i.e.~the combination of all exposures from the new 2008/2009 epoch (see \Tref{obs1}) -- for all the fitted transitions in the left, central and right regions, respectively. For transitions not covered by the 0809 sub-spectrum (see \Tref{resolving}), the 2000 sub-spectrum is plotted. 

Due to the high resolution and (mainly) very high S/N of the spectrum, plus the complexity and breadth of the absorption system, a very large number of components, 106, was required to fit the statistically significant velocity structure in all the transitions. To our knowledge, this is a much larger number than has been used to fit any metal absorption system before. Fitting such a large number of components was extremely challenging and time-consuming, even though we fitted the three different regions separately. This raises the question of how robustly we can determine the velocity structure and how to be sure that all statistically significant structure in the profile is fitted.

\begin{figure}
  \centering
    \includegraphics[width=0.49\textwidth,natwidth=610,natheight=642]{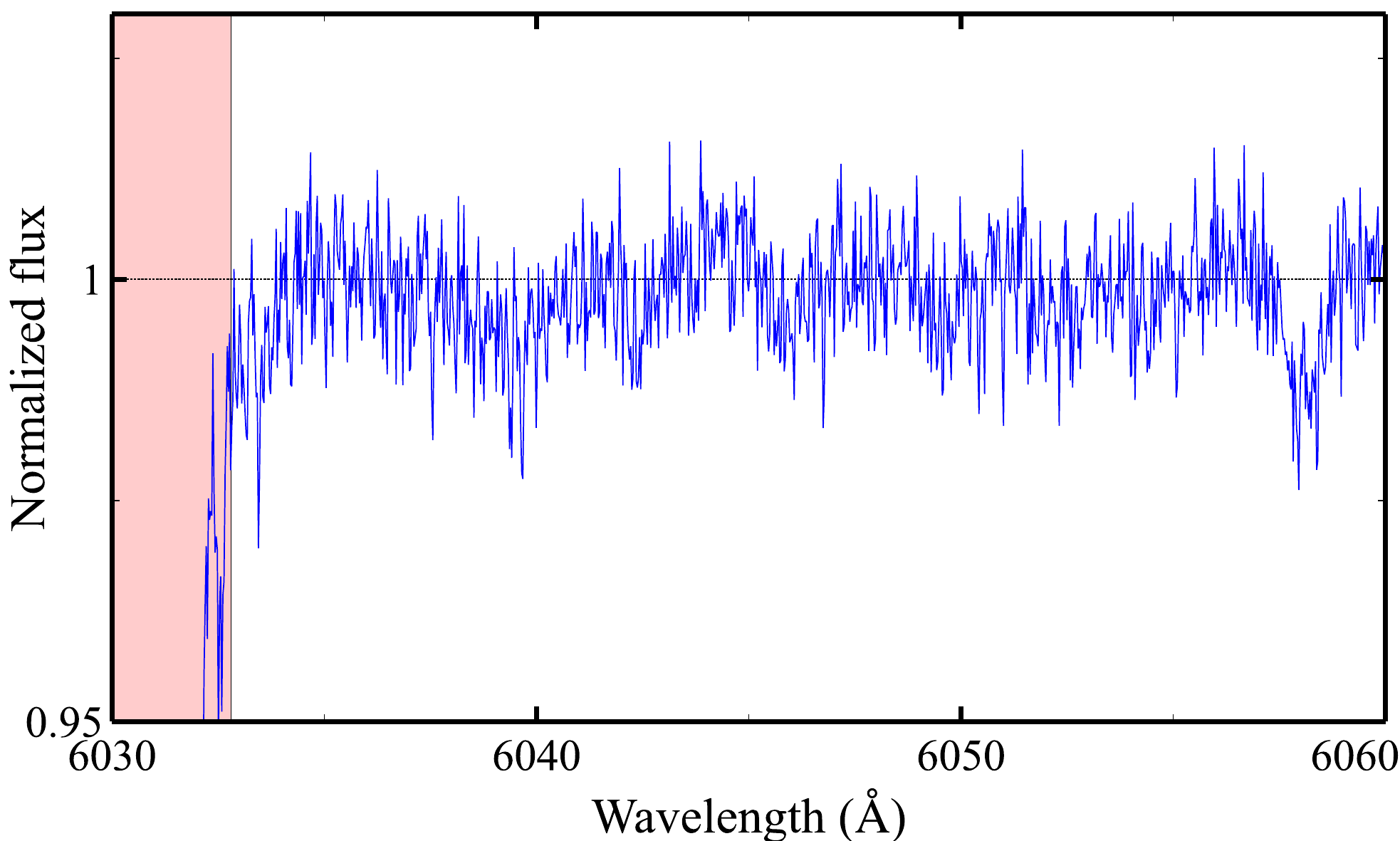}
  \caption{Example of the fringing effect in the red part of the combined 0809 sub-spectrum. Red shading represents the reddest half of the right fitting region of the \ion{Mg}{ii}\,2803 transition. Many fringes are evident as significant, low-level (note the zoomed vertical axis range) variations in the unabsorbed continuum flux over $\sim$20--100 pixel ranges. These fringes are likely to be present in the fitting regions of the \ion{Mg}{i/ii} transitions as well. Fringing with similar magnitude, but uncorrelated structure, is evident in all sub-spectra.}\label{fringing}
\end{figure}

The difficulty in establishing the preferred velocity structure was compounded by data artifacts. Figures 2--5 show the residuals between our model and the spectrum, normalized by the 1$\sigma$ error array. Note that, in some parts of some transitions, even our very complex model leaves significant residuals in several neighboring pixels (e.g.~around $+$100\,\kms\ in \ion{Mg}{ii}\,2803 in the right region, \Fref{right_a}). In many cases, especially in the redder transitions (e.g.~\ion{Mg}{i}\,2852, the \ion{Mg}{ii} doublet and the \ion{Fe}{ii} lines redwards of 2344\,\AA\ in the absorber rest-frame), these significant runs of residuals are most likely caused by low-level CCD fringing (i.e.~internal reflections within the CCD substrate). An example of the fringing effect is shown in \Fref{fringing}. In other parts of the spectrum we also find evidence for artifacts from the flux extraction, sky subtraction and flat-fielding. These effects are not normally significant in quasar spectra with $\SN\la100$ but likely dominate over the statistical noise in some places of our much higher $\SN$ spectrum. These artifacts may drive a requirement to fit more velocity components, so this complicated our approach to determining when to stop adding new components to our model profile.

Previous studies have used several different criteria to decide the best number of components to fit, e.g.~minimizing the $\chisq$ per degree of freedom, $\chisqn$, \citep[e.g.][]{2008MNRAS.384.1053M, 2013A&A...555A..68M, 2014MNRAS.445..128E}, or minimizing the Akaike information criterion \citep[e.g.][]{2012MNRAS.422.3370K}, and by ensuring that no significant structures in the residuals are common among the different fitted transitions \citep[by constructing ``composite residual spectra",][]{2010MNRAS.403.1541M}. However, the data artifacts discussed above force us to adopt a different approach to avoid fitting too many components; the $\chisq$ minimization algorithm cannot sustain an arbitrarily large number of components due to the degeneracies that develop between neighboring components. Therefore, our approach was to stop adding more velocity components when doing so reduced $\chisqn$ by less than $\approx$0.002 and when no correlated structures in the residuals were present across the fitted transitions (as visualized through the composite residual spectrum).  We discuss the implications of these data artifacts for analyses of high-\SN\ spectra from future telescopes in \Sref{artifacts}.

After preferred models were established, we unblinded the sub-spectra by removing the artificial shifts and distortions applied in the blinding process, and began the $\chisq$ minimization with $\varal$ introduced as a free parameter. We constructed five models with different starting values for $\varal$, $-5$, $-1$, 0, 1 and 5\,ppm, to assess whether the $\chisq$ minimization converged towards the same value of $\varal$. Convergence was complete in all regions, from all starting values, except for the right region with starting values $\pm$5\,ppm and in the left region for the $-5$\,ppm starting value. This is to be expected for the right region: the final value of $\varal$ for this region is close to zero with a statistical uncertainty $<$0.6\,ppm (see \Sref{results}), so the starting values of $\pm$5\,ppm are far from this solution; coupled with the complexity of the fit, and the covariance between the column densities and $b$-parameters of closely-spaced velocity components, achieving convergence in a reasonable number of iterations ($\sim$1000) is not expected. Nevertheless, these two models in the right region were converging towards the same $\varal$ value as the other models; given a longer computation time, it is evident that these extreme models would eventually converge to the same $\varal$ value as the others in the right region. However, in the left region the $-5$\,ppm model converged towards a somewhat different $\varal$ value to the models with other starting values. In this case, the final statistical error on $\varal$ (\Sref{results}) is $>$3\,ppm, so, together with the fact that all other models converged, this suggests that the multidimensional $\chisq$ space around the solution is very flat, possibly due to some degeneracy between $\varal$ and the redshift parameters of one or more velocity components. This is further explored and the systematic error associated with this is examined in \Sref{conergence_err}. Given the overall convergence of the models in most cases, we derived our estimates of $\varal$ using fits that begin with the $\varal=0$ as the starting value. We refer to them as our ``fiducial" models and to the values of $\varal$ derived from them as fiducial $\varal$ values. We use fiducial models to estimate our final $\varal$ values (by including the isotopic structures of the fitted transitions -- see \Sref{res_stat}), and for all subsequent analyses in this study.

\subsection{Implications for velocity structure from higher-resolution spectra}\label{unresolved}
Here we explore whether additional information about the velocity structure can be obtained from the higher-resolution HARPS and bHROS spectra. Specifically, can we determine whether the fiducial model, based on the lower-resolution but considerably higher-$\SN$ UVES spectra, contains too few components spaced too far apart, or perhaps too many components spaced too closely together? That is, can the HARPS and/or bHROS spectrum be used to assess whether we have over-fitted or under-fitted our UVES spectrum?

\begin{figure}
  \centering
    \includegraphics[width=0.49\textwidth,natwidth=610,natheight=642]{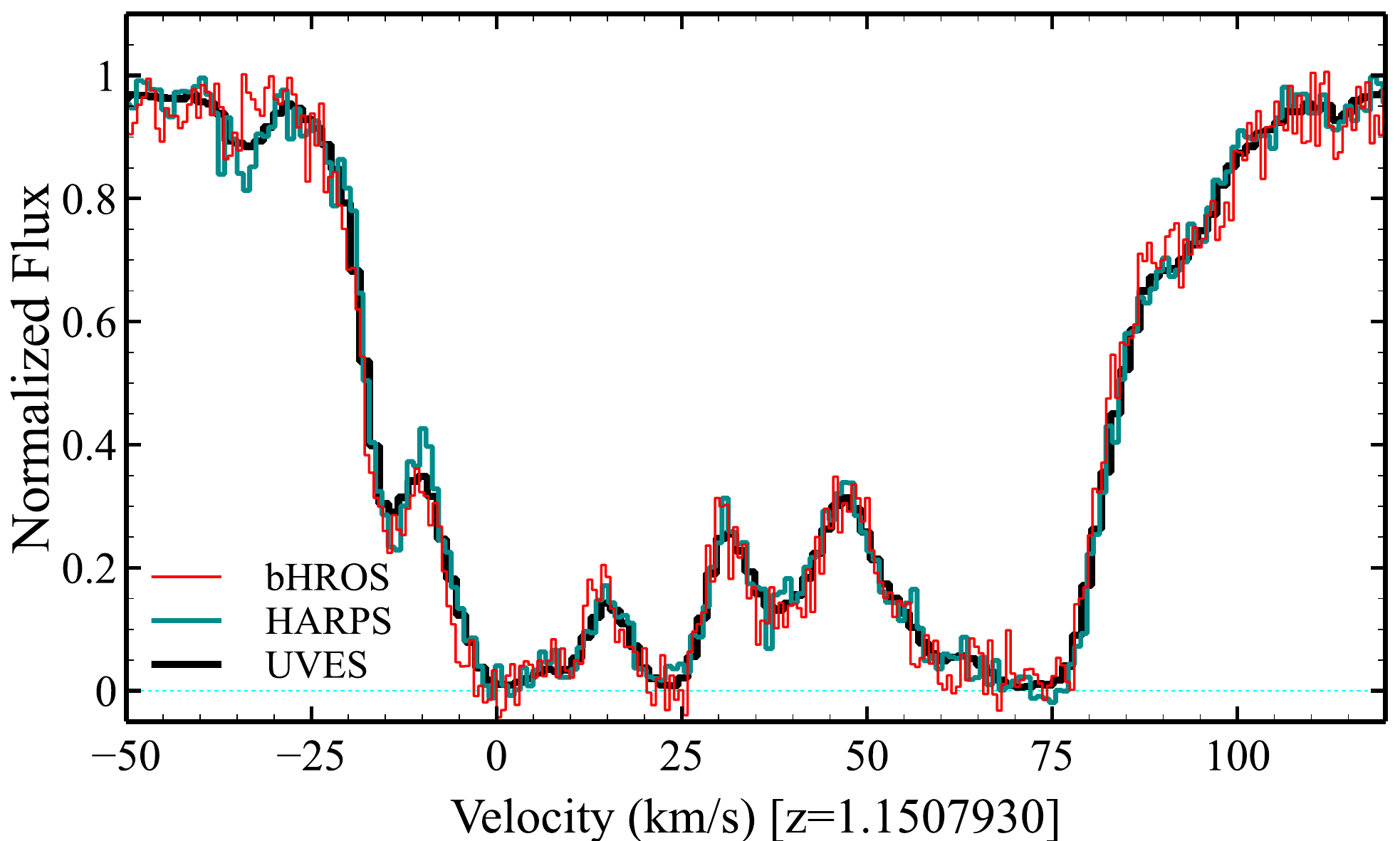}
  \caption{Comparison of the UVES, HARPS and bHROS spectra for the right region of the \ion{Fe}{II}\,2382 transition. The $\SN$ ratios per \kms\ are 200, 31 and 29 in the continuum, respectively, and it is clear that there is little evidence, if any, for additional velocity structure from the higher-resolution spectra because of their much lower $\SN$ \,per \kms.}\label{UHb}
\end{figure}

As a first step, we compared the UVES, HARPS and bHROS spectra in each transition, an example of which is shown in \Fref{UHb} for \ion{Fe}{II}\,2382. The main difficulty in comparing the spectra is already evident in this figure: the much lower $\SN$ ratio for the two highest-resolution spectra masks any possible evidence for different velocity structure, especially for the bHROS spectrum. The HARPS spectrum possibly shows some difference to the UVES spectrum at $-$35 and $-$15\,\kms. Even though these differences are small and the $\SN$ per \kms\ is $\sim$6.5 times lower for HARPS than for UVES, they may reveal information about unresolved velocity components, so we explore these differences further below. However, for this reason, it is already clear that the HARPS and bHROS spectra offer much less precise constraints on $\varal$ and no firm constraint on the reality of any individual velocity component in our fiducial model, so we do not directly include the HARPS and bHROS spectra for our $\varal$ measurement.

As a second step, we tried to assess a different, more collective property of the velocity structure: how closely velocity components are spaced relative to their widths. We constructed two simulated models, consisting of more and fewer velocity components than in the fiducial model, in a velocity range between $-15$ and 28\,\kms. The first model was constructed from our fiducial model by merging narrow velocity components into a smaller number of 4-\kms -wide (i.e. $b=4$\,\kms) velocity components. The second model was constructed from our fiducial model by separating each component broader than 1\,\kms\ into two or more, 1-\kms -wide components. We minimized $\chisq$ between these new `resolved' and `unresolved' models and the UVES spectrum, using the \ion{Fe}{II} transitions only, while holding the redshifts and b-parameters fixed and allowing the column densities to vary. We converted the instrumental profile width in both models to that of the bHROS spectrograph (i.e.\ $R=140000$) and converted the resulting flux profiles into apparent column density profiles according to
\begin{equation}
\log N_a=\log \frac{\ln (1/F)}{(f\lambda_0)\pi e^2/m_ec}\label{app_col}
\end{equation}
where $F$ is the continuum-normalized flux, $f$ and $\lambda_0$ are the oscillator strength and the transition's rest wavelength, and $m_e$ and $e$ are the mass and charge of the electron.

\begin{figure}
  \centering
    \includegraphics[width=0.49\textwidth,natwidth=610,natheight=642]{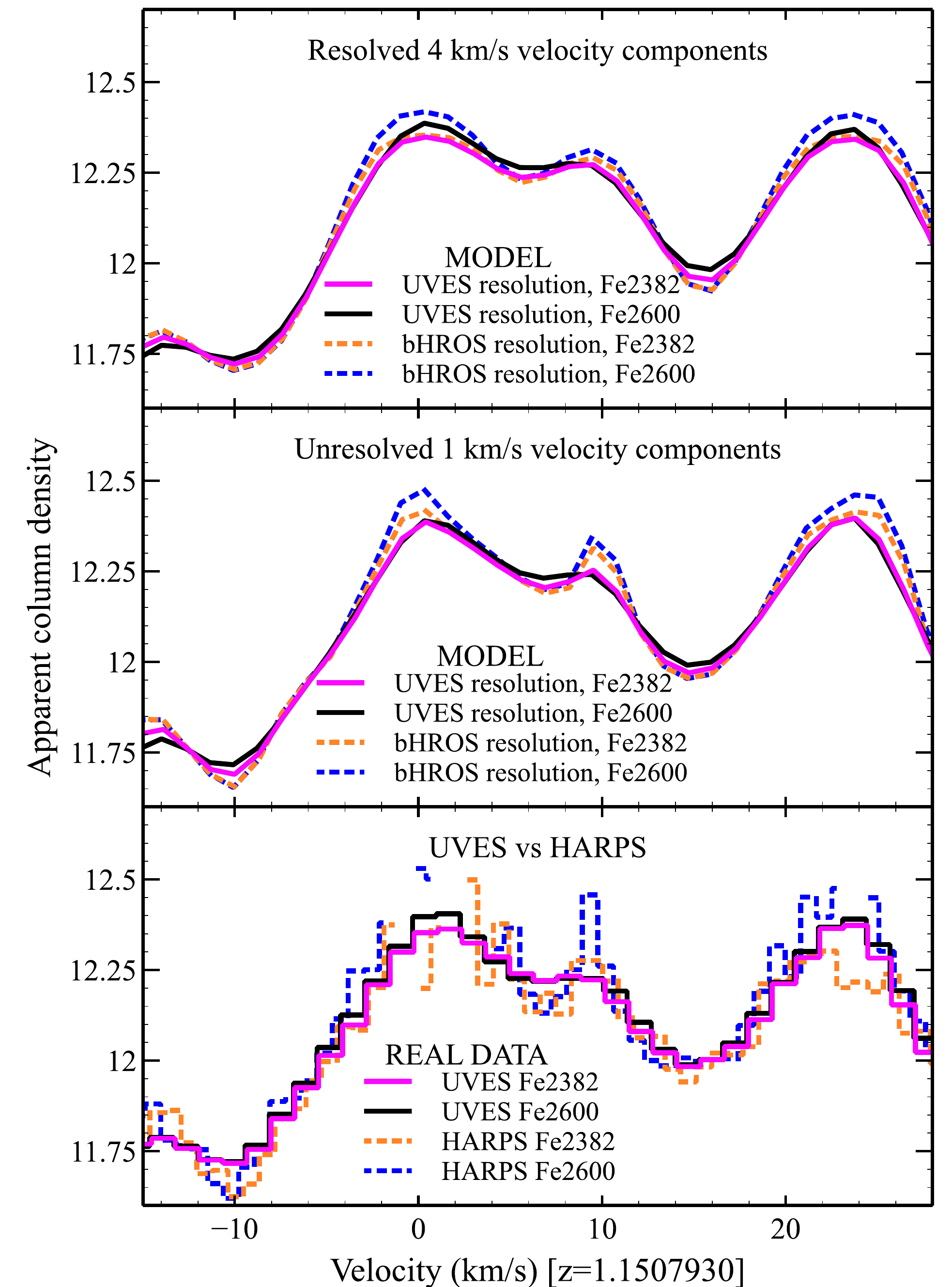}
  \caption{Apparent column density as a function of velocity in the near-saturated regions of the absorption profile for the \ion{Fe}{II}\,2382 and \ion{Fe}{II}\,2600 transitions. The upper and middle panels show two different models with the $b$-parameters of all velocity components from our fiducial model converted to 4\,\kms\ and 1\,\kms, respectively. The lower panel shows the apparent column density of the real UVES and HARPS spectra (pixels in which noise fluctuations gave negative flux values are not plotted). The profiles match very well between the \ion{Fe}{II}\,2382 and \ion{Fe}{II}\,2600 transitions at UVES resolution in the saturated regions at 0, 10 and 24 \kms (pink and black solid lines in the middle panel). However, they do not match as closely at higher, bHROS resolution, or in the case of broader velocity components which are resolved at UVES resolution. This implies that, if we have a closely-packed velocity structure in reality, with individually-unresolved components at UVES resolution, we should expect \ion{Fe}{II}\,2382 and \ion{Fe}{II}\,2600 apparent column density profiles to match well. This is what we observe in the lower panel, providing some evidence for a closely-packed velocity structure in reality.}\label{ACD}
\end{figure}

\Fref{ACD} compares the apparent column density profiles of the \ion{Fe}{II}\,2382 and 2600 transitions, for both resolved and unresolved models at both UVES and bHROS resolution. By offering a comparison of the apparent column density profiles of these transitions, both in saturated and unsaturated parts of the profile, the \ion{Fe}{ii}\,2382 and 2600 transitions give the best opportunity for understanding how tightly velocity components are packed. We find that these transitions' $N_a$ profiles match well even in saturated regions when the velocity structure comprises many closely-packed components that are not individually resolved, i.e.~the UVES-resolution model (pink and black solid lines) in the middle panel of \Fref{ACD}. There is no such match when all velocity components are resolved,  either if we increase the resolving power of the spectrograph (orange and blue dashed lines in the middle panel of \Fref{ACD}), or decrease the number of components and increase their widths (pink and black solid lines in the upper panel of \Fref{ACD}).


We compare this model prediction with the real apparent column density profiles of \ion{Fe}{II}\,2382 and 2600 transitions in the UVES and HARPS spectra in the lower panel of \Fref{ACD}. We use the HARPS spectrum here, even though the bHROS spectrum has somewhat higher resolution, because the $\SN$ of the bHROS spectrum is substantially lower than for the HARPS spectrum. The apparent column density profiles match well in the saturated regions of the UVES spectrum (pink and black solid lines in the lower panel of \Fref{ACD}), similar to the UVES-resolution models in the middle panel. This may provide some evidence that the real velocity structure has even more closely-spaced, narrow velocity structure than our fiducial model. As such, our approach of fitting a larger number of velocity components is supported to some extent by \Fref{ACD}.

Unfortunately, from comparison of the same two transitions in the HARPS spectrum (orange and blue dashed lines in the lower panel of \Fref{ACD}) we are unable to reveal how well the apparent column density profiles match. The main reasons for this are the low $\SN$ of the HARPS spectrum, which is immediately evident in \Fref{ACD}, and that this causes many pixels in highly saturated regions to have flux values lower than zero (missing in \Fref{ACD}). These cannot be converted to apparent column density with \Eref{app_col}. However, even if we had a high $\SN$\,per\,\kms\ spectrum at higher resolution, and the real velocity structure comprised very closely-packed, narrow velocity components, we would still not be able to resolve them individually -- they would be too closely packed to produce anything but smooth features in the spectrum. Therefore, increasing the resolution would only slightly increase the amount of `centroiding information' and the precision with which we can measure $\varal$ is proportional to this. In other words, even with a similar $\SN$\,per\,\kms\ as our UVES spectrum, observing at higher resolving power would not substantially increase the precision on $\varal$. 

\section{Results}\label{results}
\subsection{Results from {\sc vpfit} and statistical errors}\label{res_stat}
The best-fitting value of $\varal$, its 1$\sigma$ statistical error and the $\chisq$ per degree of freedom, $\chisqn$, for each of the three regions are shown in \Tref{results}. These values were derived by minimizing $\chisqn$ in {\sc vpfit} from the fiducial model for each region. The statistical errors are determined from the diagonal terms of the final parameter covariance matrix and, for a given profile model, derived only from the statistical flux noise in the spectra. Systematic errors are discussed in \Sref{systematic_err} below. \Sref{comb_err} explains how we combined the values from the individual regions to form a combined $\varal$ estimate and error budget for the entire absorber.

The final results in \Tref{results} were derived using the laboratory wavelength information recommended by \citep{2014MNRAS.438..388M}, including the isotopic structures (where known from measurements or theoretical calculations). We assume terrestrial isotopic abundances, even though we cannot be sure that the same isotopic abundance pattern exists in this absorbing cloud at $z=1.15$. We discuss the impact of different isotopic abundances in the cloud, especially for \ion{Mg}, in \Sref{without_Mg}. The final fitted parameters and their uncertainties from {\sc vpfit} are provided in electronic format in \citet{srdan_kotus_2016_51715}.

The ESO UVES Common Pipeline Library, which we used to reduce the spectra, fails to correctly estimate the flux errors in near or fully saturated parts of the spectra \citep{2012MNRAS.422.3370K}. Therefore, we used the modified version of {\sc vpfit} \citep{2012MNRAS.422.3370K} which increases the error array appropriately in such regions for the correct estimate of the $\chisqn$ in the \Tref{results}.

It is evident in \Tref{results} that the right region measurement is by-far the most precise among the three measurements. This is mainly because this is the most complex, highest column-density part of the absorption system: there are more sharp, deep features from which constraints on the relative velocity between the transitions can be derived. While more transitions are also utilized in this region (i.e.~those of \ion{Ni}{ii}, \ion{Mn}{ii}, \ion{Cr}{ii} and \ion{Zn}{ii}), they are all very weak and consequently contribute little to the statistical precision on $\varal$ in this region (see analysis in \Sref{just_Fe} in which these transitions are removed). The complexity of the absorption in the right region, combined with the data artifacts discussed in \Sref{fitt}, especially fringing effects, lead to a higher $\chisq$ in that region. Indeed, among the 25 transitions fitted in this region, $\chisqn$ is largest for the \ion{Mg}{ii} transitions where the fringing is most important. However, the larger $\chisq$ for the \ion{Mg}{ii} transitions may also be due to the difficulty in fitting these much stronger and somewhat saturated transitions; there may exist weak velocity components that are not fitted and which contribute negligibly to absorption in other species (e.g.~\ion{Fe}{ii}).

The statistical uncertainties on $\varal$ range between 0.592 and 3.381\,ppm across the three regions. The latter is comparable with the most precise previous measurements of $\varal$ in individual absorption systems (Q04; \citeauthor{2007A&A...466.1077L}, \citeyear{2007A&A...466.1077L}; M08b; \citeauthor{2011A&A...529A..28A}, \citeyear{2011A&A...529A..28A}; \citeauthor{2013A&A...555A..68M}, \citeyear{2013A&A...555A..68M}; \citeauthor{2014MNRAS.445..128E}, \citeyear{2014MNRAS.445..128E}) and the former is comparable with, though smaller than, the statistical error in the weighted mean $\varal$ values reported for the large samples of absorbers studied in \citep{2004LNP...648..131M, 2012MNRAS.422.3370K}.

\begin{table}
\caption{Final measurements of $\varal$ for the three fitting regions and the combined value with $1\sigma$ statistical and systematic uncertainties. $\chisqn$ is given in the last column. We provide additional significant figures here to allow reproducibility of our calculations only.}\label{results}
\centering
  \begin{tabular}{@{} l c c  }
  \hline   
  Region   &   $\varal$ [ppm]   &     $\chisqn$     \\   
  \hline         
   Left     &  $-6.153\pm3.381_{\rm stat}\pm2.055_{\rm sys}$   &    $1.086$   \\
   Central  &  $-4.692\pm1.743_{\rm stat}\pm1.612_{\rm sys}$   &    $1.171$   \\
   Right    &  $-0.843\pm0.592_{\rm stat}\pm0.554_{\rm sys}$   &    $1.262$   \\
   Combined &  $-1.422\pm0.553_{\rm stat}\pm0.645_{\rm sys}$   &              \\
  \hline 
  \end{tabular}
  \end{table}

Using only the statistical errors, the $\varal$ values in the three regions are marginally inconsistent with each other: the values from the left and central regions are 1.5 and 2.1$\sigma$ from that of the right region. The $\chisq$ around the weighed mean value, $-1.37\pm0.55$\,ppm, is 6.42. Formally, this value (or higher) has a 4\,per cent chance of occurring randomly, but with just 3 values to compare, this estimate is not reliable. More importantly, systematic effects have not yet been included, and we discuss these below. Some of those effects are correlated between the absorbers, so combining them and discussing their consistency requires further analysis in \Sref{comb_err}.

\subsection{Systematic error estimates}\label{systematic_err}
Given that we study only a single absorption system in this study, an important goal is to identify all systematic effects that can influence our results and to quantify them. {\sc vpfit} calculates statistical errors on $\varal$ very robustly \citep{2009MmSAI..80..864K}, but it is unable to provide any information about the possible presence of systematic errors. In this Section we estimate the effect on $\varal$ of the four most important systematic errors: long-range wavelength distortions, intra-order distortions, redispersion of exposures onto the common wavelength scale, and inaccuracies in the modeled velocity structure. These are discussed in the following subsections, and \Sref{conergence_err} discusses the additional systematic error in the left region caused by a relatively flat $\chisq$ space. The systematic uncertainties are summarized in \Tref{systematics}, which also shows the total systematic error for each region (calculated as the quadrature sum of individual systematic uncertainties for each region). Several other consistency checks, which limit the cumulative effect of these main systematic errors, are conducted in \Sref{consistency_checks} together with other possible effects, such as isotopic abundance variations and real velocity shifts between ionic species. However, those consistency checks do not modify our systematic error budget.

\subsubsection{Long-range wavelength distortions}\label{long-range_sys}
In our analysis we have compared UVES sub-spectra with the HARPS spectrum of the same object to correct for the long-range wavelength distortions before measuring $\varal$. Here we use the uncertainties in the distortion slopes derived in that comparison (using the DC method), which are shown in \Tref{slopes}, to estimate the systematic uncertainty implied for $\varal$.

For each sub-spectrum, we imposed an additional distortion slope on each of its constituent exposures, equal to the slope uncertainty, +1$\sigma$, and recombined the sub-spectrum again in {\sc uves\_popler}. We also created another sub-spectrum with a $-$1$\sigma$ correction applied. {\sc vpfit} is then used to optimize the fiducial model (separately) on the two new sets of sub-spectra (+1$\sigma$ and $-$1$\sigma$) to determine how much $\varal$ differs from the fiducial value in each of the 3 fitting regions. The long-range distortion systematic error is then computed as the mean of the two estimates in each region. The systematic errors calculated from this analysis are $\pm$1.22, $\pm$1.05 and $\pm$0.51\,ppm, for the left, central and right regions, respectively, and these are shown in \Tref{systematics}. 

It is also interesting to determine how large the systematic effect from the long-range distortions would have been if we had not corrected them in the UVES spectra. To assess this we remeasured $\varal$ in the right region after recombining the sub-spectra without correcting them for the distortions found in \Sref{DC_HARPS}. After optimizing the model in {\sc vpfit} we found that $\varal$ was 2.1\,ppm lower than our fiducial value for the right region. In \Sref{DC_HARPS} we estimated that the long-range distortions we find (i.e.~a slope of $\approx$200\,\ms\,per 1000\,\AA) could affect a $\varal$ measurement by up to $\sim$10\,ppm if, as an example, only the \ion{Mg}{ii}\,2796 and \ion{Fe}{ii}\,2382 transitions were considered. Comparing that estimate with our measured value of 2.1\,ppm, it is clear that the constraints on $\varal$ in our absorption system are not solely dominated by the comparison of Mg and \ion{Fe}{ii} transitions redwards of (rest-frame) 2340\,\AA\ like, for example, many of the absorbers in the large samples of \citet{2004LNP...648..131M} and \citet{2012MNRAS.422.3370K}, particularly those at redshifts $\zab<1.6$. As we show below in Sections \ref{without_blue} and \ref{just_Fe}, the comparison between those red \ion{Fe}{ii} transitions and \ion{Fe}{ii}\,1608 is more important in our absorber. A simple estimate of the expected effect on $\varal$ considering only the \ion{Fe}{ii}\,1608 and $\lambda$2382 transitions, similar to the previous one using \ion{Mg}{ii}\,2796 and \ion{Fe}{ii}\,2382, demonstrates that a 2\,ppm systematic error is expected for our absorber. This estimate takes into account the fact that these transitions fall in the two different arms of UVES and assumes that the long-range distortions are separate in the two arms but approximately aligned (with zero velocity difference) near the centres of their respective wavelength ranges.

\subsubsection{Intra-order wavelength distortions}\label{intra-order_sys}
Intra-order wavelength distortions are those on scales of, and which tend to be repeated across, individual echelle orders. They were identified for the first time in \citet{2010ApJ...708..158G}. \citet{2015MNRAS.447..446W}, using many supercalibration spectra of asteroids and solar twins from UVES, found that the intra-order distortions were typically very similar across all echelle orders in a single exposure, with little change between exposures over several-day timescales. They found their typical amplitude to be $\Delta v\sim100$\,\ms. Previous studies \citep[][]{2010MNRAS.403.1541M, 2013A&A...555A..68M, 2014MNRAS.445..128E, 2014ApJ...782...10B} have estimated the systematic error on varying-constant measurements by modelling the intra-order distortion as a simple “saw-tooth” function, with the velocity shift varying linearly from a peak, $\Delta v$, at the centre of each echelle order to $-\Delta v$ at the order edges. The value of $\Delta v$ was estimated from the intra-order distortions observed in supercalibrations observed as close as possible to the quasar exposures. 

In this work, we follow the same modelling approach, but we are unable to estimate the intra-order distortion amplitude, $\Delta v$, because we do not have complementary supercalibration exposures. Therefore, we assume typical values for the distortion amplitude for UVES found in previous supercalibration studies \citep{2015MNRAS.447..446W}, $\Delta v = 100$\,\ms. To simplify the analysis we assume the same distortion in all of our exposures. We introduce this distortion for each echelle order, recombine the spectra in {\sc uves\_popler} and use these adjusted spectra to measure $\varal$. The difference between the fiducial $\varal$ and the $\varal$ measured in this analysis provides the intra-order systematic error estimate: $\pm$0.12, $\pm$0.37 and $\pm$0.16\,ppm, for left, central and right regions, respectively.

One might at first expect this effect to be larger than we measure it to be because we study a single absorber and some transitions of interest should be at the edges of echelle orders and highly susceptible to this distortion. However, this is not the case because, as we described in \Sref{UVES_ORC}, we have rejected pixels at the edges of echelle orders if they overlapped with our transitions of interest. The effect is also reduced by using a large number of transitions, as we do here, with the main constraints on $\varal$ contributed by the 9 \ion{Mg}{i/ii} and \ion{Fe}{ii} transitions shown in \Fref{right_a}. Furthermore, the intra-order distortions in the combined spectrum should average out in the real spectra to some extent if they vary to some degree over the many years of exposures. Therefore, our intra-order systematic error estimates may be even somewhat conservative. 

\begin{table}
\caption{Contribution of individual systematic errors to the systematic error budget on $\varal$ in each region and the total systematic error estimate in each region, in units of ppm. The systematic error for the entire absorption system is also presented, together with the contributions from the individual effects, as described in \Sref{comb_err}. We provide additional significant figures here to allow reproducibility of our calculations only.}\label{systematics}
\centering
  \begin{tabular}{@{} l c c c c}   
  \hline
Systematic effect & \multicolumn{3}{c}{Region} & Entire  \\
                                                & Left                                    &  Central                              & Right        & system  \\
  \hline
 Long-range distortions         & 1.220 &  1.047  & 0.512 & 0.593 \\
 Intra-order distortions        & 0.119 &  0.372  & 0.159 & 0.185 \\ 
 Redispersion                     & 0.510 &  0.369  & 0.041 & 0.070 \\
 Velocity structure             & 0.393 &  1.108  & 0.131 & 0.160 \\
 Convergence                    & 1.518 &        &      & 0.069     \\
 Quadrature sum                       & 2.055 &  1.612  & 0.554 &  0.645    \\
  \hline 
  \end{tabular}
  \end{table}

\subsubsection{Spectral redispersion effects}\label{sys_disp}
To construct our sub-spectra we used a dispersion of $1.3$\,\kms\,$\text{pixel}^{-1}$ for 1$\times$1-binned exposures and $2.5$\,\kms\,$\text{pixel}^{-1}$ for 2$\times$1-binned exposures. The dispersion defines the binning grid and, when each exposure is redispersed onto the common wavelength grid of the final spectrum, the ``phasing" of the binning grid changes. This causes slight shifts in the fitted line centroid position. Furthermore, the rebinning of individual exposures onto the same wavelength grid introduces correlations between the fluxes and flux uncertainty estimates of neighboring pixels. 

To estimate the systematic error caused by these two effects we re-combine the sub-spectra in {\sc uves\_popler} with dispersions differing from the fiducial sub-spectra by $\pm$0.05 and $+0.1$\,\kms\ for 1$\times$1-binned exposures and $\pm$0.1 and $+0.2$\,\kms\ for 2$\times$1-binned exposures. We then measure $\varal$ by rerunning {\sc vpfit} with these newly created sub-spectra and our fiducial models as starting points. We estimate the systematic errors caused by spectral redispersion effects as the mean of the difference between the fiducial value and these estimates of $\varal$. The results are $\pm$0.51, $\pm$0.37 and $\pm$0.04\,ppm, for the left, central and right regions, respectively. 

\subsubsection{Velocity structure modelling errors}\label{sys_v_str}
Our fitted velocity structure in each region consists of a large number of velocity components. While we thoroughly explored many different possible velocity structures, and the normalized residuals in our fiducial fits, shown in Figs.~2--5, show no strong evidence of additional, unfitted structure, it remains likely that different velocity structures, with slightly more or fewer components, may provide equally satisfactory fits. It is also possible that some velocity components used in our fiducial models were required by the data artifacts that, at the very high $\SN$ of our spectra, cause statistically significant structure in some transitions (albeit uncorrelated from transition to transition). That is, overall, we cannot be sure that our fitted velocity structure is unique or correct, even though it provides a statistically acceptable fit, so we explored many alternative velocity structures to estimate the systematic uncertainty related to these effects. 

To measure possible uncertainties in the velocity structure we trailed several deviations from our fiducial model by removing several velocity components one by one, while still trying to achieve reasonable residuals between the sub-spectra and the new model. We have two approaches to decide which components to remove: (1) if the component is very close to another component and/or (2) if it is missing from some of the weaker species. We measure $\varal$ by minimizing $\chisq$ in {\sc vpfit} and calculate the velocity structure systematic error for each case as the difference between our fiducial $\varal$ value and the value after removal of the component. We obtain the systematic error by averaging these differences among different models. 

We have excluded several cases from this analysis if they did not pass certain criteria. Firstly, we required a reasonably low $\chisqn$, i.e.~similar to that associated with our fiducial fit. This was necessary to avoid accepting fits that would not have passed our original criteria for selecting our fiducial model. Secondly, we did not accept models which left substantial deviations from zero in the normalized residual spectra, particularly when those were clearly (anti-)correlated across several transitions. Usually these cases also showed higher $\chisqn$. Finally, we did not accept models from which strong velocity components (i.e.~those with high optical depths in \ion{Mg}{i/ii} and \ion{Fe}{ii}) were removed by {\sc vpfit} during the $\chisq$ minimization process.

For each trial velocity structure, we detail the components that were removed and which models remained `viable' (in terms of the criteria above) in subsequent paragraphs.

\textbf{\textit{Left region:}}
Firstly, we removed one of the two components in the strongest `feature' (i.e.~region of the spectrum between its local maxima) at $-$535\,\kms. Removing either of the two components produces similar results. The only requirement is to remove the same component from different species. After its removal, the component in the wing of this feature, which is relatively strong in \ion{Mg}{ii}, drops from \ion{Fe}{ii} during the {\sc vpfit} $\chisq$ minimization process. However, after the minimization was complete, we did not observe significant or correlated residuals and the $\chisqn$ was only larger than that of our fiducial model by 0.006. Therefore, we include this result for the systematic uncertainty estimate. 

Secondly, if we remove the leftmost component in the feature at $-$415\,\kms, the rightmost component in this feature drops from \ion{Fe}{ii} during the $\chisq$ minimization process. In this case the $\chisqn$ increases by only 0.003 and we do not see any strong correlations between residuals. Therefore, we include this result for the estimate of the uncertainty.

Thirdly, we removed one of the components in the feature at $-$400\,\kms. After $\chisq$ minimization the component from \ion{Mg}{i} in the feature at $-$415\,\kms\ dropped during the $\chisq$ minimization process and the other component from this same feature shifted towards the initial velocity of the removed component. This indicates that the removed component is statistically important for the fit. Because of this drop and shift, this case is very similar to the previous case. On the other hand, the $\varal$ estimate is significantly different from the previous case, which appears to be due to the convergence difficulties described in \Sref{conergence_err}. Furthermore, the residuals do not increase significantly and are not correlated between transitions, and $\chisqn$ increases by only 0.003. Therefore, we include this result for the estimate of the uncertainty.  

Finally, we removed the middle component from the feature at $-$495\,\kms, which resulted in the same $\chisqn$ as for the fiducial fit, without any components dropped. This result is also included in the uncertainty estimate.

When we combine these four systematic error estimates by averaging them, we measure the velocity structure modelling error in the left region to be $\pm$0.39\,ppm. 

\textbf{\textit{Central region:}}
Firstly, we removed one of the two components in the strongest feature at $-$235\,\kms. During the $\chisq$ minimization procedure several components dropped and $\chisqn$ increased by more than 1 from its fiducial value; this was also reflected in a large increase and strong correlation in the residuals. Therefore, we exclude the result of this trial from the estimate of the systematic uncertainty.

Secondly, we removed one of the components in the $-$135\,\kms\ feature. The $\chisq$ minimization in this trial produced a $\chisqn$ which was 0.03 larger than the fiducial value. However, the residuals did not increase substantially and no strong correlations were apparent across many transitions, so we include this $\varal$ value in the uncertainty estimate.  

Thirdly, we removed the middle component at $-$185\,\kms\ in the left wing of the feature at $-$180\,\kms. In this case one unimportant component dropped from \ion{Al}{iii}. The $\chisqn$ even decreased in this case in comparison to our fiducial $\chisqn$ and the residuals did not increase noticeably or become more correlated between transitions. Therefore, we include this value in the uncertainty estimate.

Finally, we removed the left component in the weak feature at $-$300\,\kms. The $\chisqn$ was the same in this case and there was no apparent change in the residuals, so we include this value in the uncertainty estimate.

We use the average of these measurements, $\pm$1.11\,ppm, as the velocity structure modelling error in the central region. 

\textbf{\textit{Right region:}}
The right region has two features, at 0 and 25\,\kms, where several components are fitted in the fiducial model. In both features, removing any of these components from the model produces much larger $\chisqn$ values than our fiducial model. This was expected because fitting these features required many different velocity structures to be trialed while establishing our fiducial model. Therefore, we do not include these trial fits with one or more of those components removed in our modelling uncertainty estimate for the right region. 

We investigated the effect of removing other components located at $-18$, 12, 43 and 78\,\kms. All of them produced $\chisqn$ values close enough to the fiducial fit. However, for the 12\,\kms\ case, several components in other parts of the region dropped from weak ionic species during the $\chisq$ minimization process, and for the 78\,\kms\ case the component at 77\,\kms\ was dropped from \ion{Mg}{ii}. Regardless, we decided to keep these results, mainly because we observed no substantial changes in the residuals. 

After averaging the estimates of the deviations in $\varal$ we find a systematic uncertainty associated with velocity structure modelling for the right region of $\pm$0.13\,ppm.

\subsubsection{Convergence error}\label{conergence_err}

Due to the very complex velocity structure in all regions, our models have a large number of velocity components and free parameters. {\sc vpfit} has been used previously to fit systems with far fewer components. However, to test whether {\sc vpfit} minimizes $\chisqn$ correctly for our much more complex model, we tested whether the models converge towards the same $\varal$ value when we start the fitting procedure from different values of $\varal$ in \Sref{fitt}. The relevant models converged in the central and right regions to within 0.2 and 0.08\,ppm of our fiducial values, so we do not incorporate an additional term in the systematic error budget for these regions.

However, as described in \Sref{fitt}, a convergence problem became evident in the left region, where the model that started from $-$5\,ppm converged towards $-$7.2\,ppm and all other models converged to within 0.4\,ppm of $-$4.2\,ppm. This implies that variation in $\chisq$ with $\varal$ in the vicinity of the best solution is almost entirely compensated by variations in other parameters, a degeneracy which seems to be caused, at least in part, by the 3 strongest components in the strongest feature in this region at $-$535\,\kms. Indeed, when conducting other consistency checks and tests for other systematic errors, $\varal$ often converged to values close to one of these two values, $-7.2$ and $-4.2$\,ppm. 

Our fiducial value in the left region is therefore the mid-point between these two extreme values, and we include an additional systematic error term -- a convergence error -- of $\pm$1.52\,ppm for this region. When deriving the other terms in the systematic error budget for this region, we measured the deviation of $\varal$ from the closest of these two extremes. It is therefore possible that the other elements of the systematic error budget for this region also include some degree of this convergence error and may therefore be somewhat overestimated. 

This highlights a particular problem with fitting very high $\SN$ spectra with multi-component Voigt profile models and may indicate that we are reaching a limitation of this approach.        

\subsection{Consistency checks and astrophysical systematic effects}\label{consistency_checks}

Here we conduct several consistency checks and discuss the potential for astrophysical systematic effects in the absorption system. We find that we do not need to include the results of the former into the systematic error budget. However, the extent to which the latter have affected our results is unclear and difficult to incorporate into the formal systematic error budget.

\subsubsection{Testing the possibility of different long-range wavelength distortions in the blue and red settings}\label{without_blue}

When correcting for the long-range wavelength distortions in \Sref{DC_HARPS} we assumed that the long-range distortions in the red and blue settings were the same in each sub-spectrum. While \citet{2015MNRAS.447..446W} found this to be a good approximation for most supercalibration exposures from UVES, it was not always true. Therefore, this assumption might not be true for some of the sub-spectra used in this study.

We test this by measuring $\varal$ using only transitions that fall in the wavelength range of the 580-nm setting. This includes all \ion{Mn}{ii}, \ion{Mg}{i/ii} and \ion{Fe}{ii} transitions, excluding \ion{Fe}{ii}\,1608 and \ion{Mg}{i}\,2026. This ensures that we include only transitions for which we directly measured the long-range distortion from the DC method. The $\chisq$ minimization procedure was started from our fiducial fit in the right region but with $\varal$ reset to 0\,ppm. This region provides the most valuable test because it yields the tightest constraint on $\varal$. The resulting $\varal$ value was $-2.62\pm1.89_{\rm stat}$\,ppm for this fit that includes only transitions that fall in the red arm. 

The statistical error bar for this measurement is $\sim$3 times higher than our fiducial error bar (0.59\,ppm). Such a large increase indicates that strong constraints on $\varal$ come from combining the transitions in the blue and red arms, not just from those in one arm alone. This is reasonable to expect because the \ion{Fe}{ii}\,1608 transition provides a strongly contrasting $q$ coefficient to those of the \ion{Fe}{ii} transitions falling in the red arm. This test shows the importance of this specific transition for the $\varal$ measurement. On the other hand, the difference between $\varal$ from this test and our fiducial measurement ($-0.92$\,ppm) is only a factor of 1.3 larger than the difference between their error bars. Clearly, the measurements are correlated; but in simple terms this indicates that a deviation of this magnitude is expected. We therefore conclude that they are not significantly different. This indicates that there is no evidence that there are substantial differences in the long-range distortions for the blue and red arms.       

\subsubsection{Further, unmodeled velocity structure}\label{just_Fe}

Some previous studies have claimed that the Many Multiplet method is not appropriate because there is a possibility for different kinematics, and therefore velocity structure differences, between different species \citep[e.g.][]{2005A&A...434..827L}. However, this would imply that different ions are physically separated in the individual clouds and have their own distinct parameters (e.g.\ temperature and turbulent motion). We argue that this is unlikely and, rather, that it is important to distinguish such ``kinematic'' differences between species from another, much more likely possibility: significant differences in relative column densities in neighboring velocity components in different species. If, for example, there exists two closely-spaced velocity components which are mistakenly fitted as a single component, and the relative column densities in these two components vary substantially between different species, then a spurious velocity shift will be measured between the species. One can generalize that example to cases where the physical conditions vary somewhat across individual clouds. This is another important motivation for our approach to fitting the absorption system with as many velocity components as required to account for all statistically significant structure. Of course, it is possible that we are still under-fitting the absorber to some extent, as discussed in \Sref{fitt}.

To check the potential effect this may have had on $\varal$, we repeated the $\chisq$ minimization process using only the \ion{Fe}{ii} transitions, excluding all other transitions in the right region. The $\chisq$ minimization procedure was started from our fiducial fit in the right region but with $\varal$ reset to 0\,ppm. The resulting $\varal$ value was $-0.62\pm0.61_{\rm stat}$\,ppm for this Fe-only fit. This is only slightly larger than our fiducial value in the right region ($-0.92\pm0.59_{\rm stat}$\,ppm) and we conclude that this effect is not important. This comparison is also sensitive to any residual long-range distortions because the \ion{Fe}{ii}\,1608 transition's $q$ coefficient has an opposite sign to that of the redder \ion{Fe}{ii} transitions. Therefore, the result of this test provides additional confidence that the long-range distortions have been removed adequately and without significant residuals. It is also interesting to note that the \ion{Fe}{ii} velocity structure remains the same after removing the constraints on the velocity structure from other species. No components are excluded by {\sc vpfit} from the fit during the $\chisq$ minimization process. This confirms that the large number of velocity components is necessary even to fit only \ion{Fe}{ii}.

\subsubsection{Measuring velocity shifts between strong transitions}

Measuring $\varal$ is, at the most basic level, the measurement of a pattern of velocity shifts between several transitions. However, if systematic effects introduce additional velocity shifts between the transitions, this can shift the measured $\varal$ value. The intra-order distortions, for example, should typically introduce shifts of $\sim$100\,\ms\ between transitions and, because they seem to follow the same pattern for all orders, the shifts will depend on the transitions' relative positions along their respective echelle orders. A generic test for such systematic effects is to measure the velocity shifts between transitions using our fit while assuming $\varal$ is zero.

We conducted this test in the right region using only the strongest transitions, those of Mg and Fe, because the weak transitions (e.g.\ of Mn, Ni etc.) provide velocity shift measurements that are not precise enough to be useful. A single velocity shift parameter was introduced for each transition in {\sc vpfit} and assumed to be the same for all sub-spectra. The \ion{Mg}{ii}\,2852 transition is taken as the reference (its velocity shift is fixed at zero). The $\chisq$ minimization procedure was started from our fiducial fit in the right region.

\begin{figure}
  \centering
    \includegraphics[width=0.49\textwidth,natwidth=610,natheight=642]{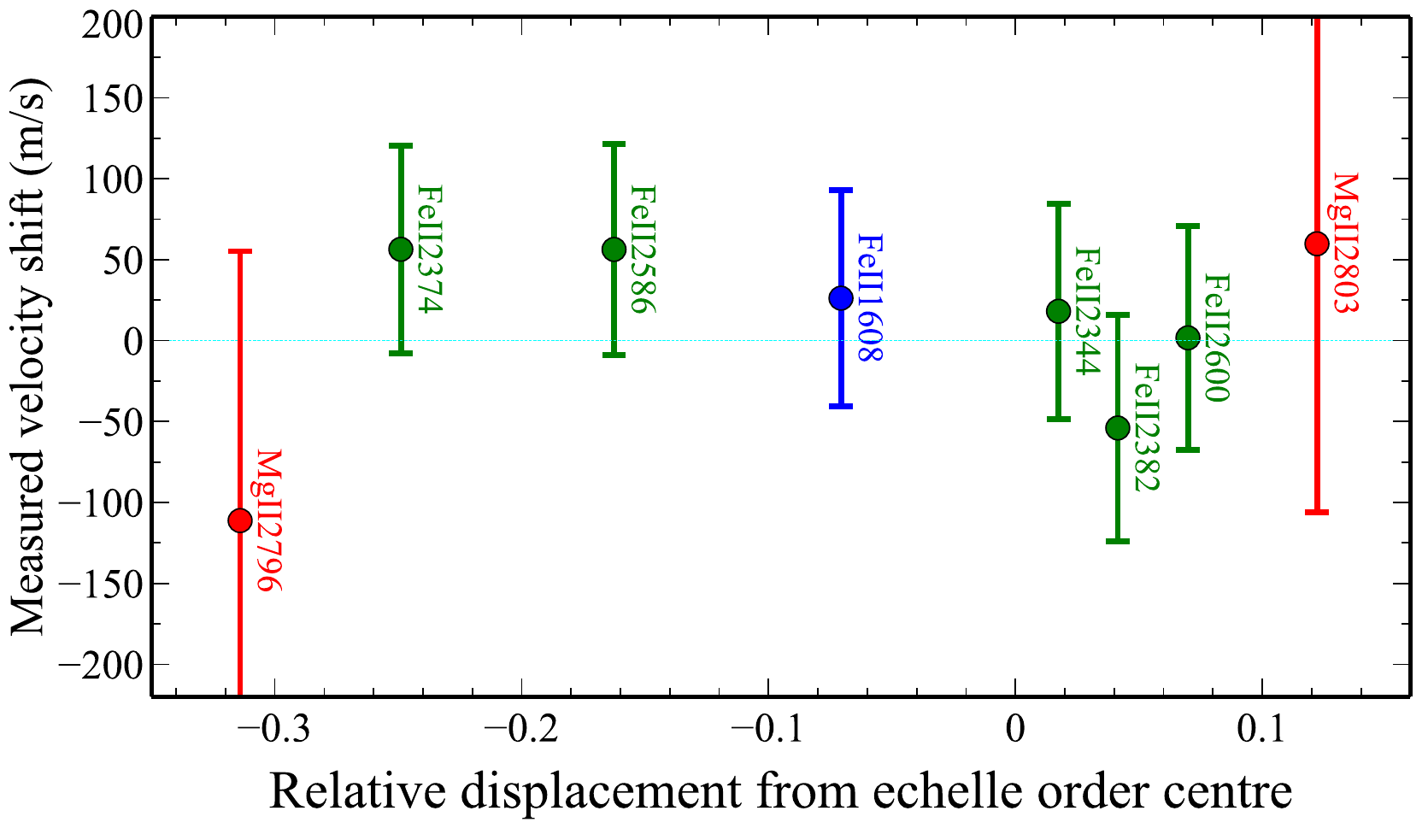}
  \caption{Measured velocity shift as a function of relative position along the echelle order for the strongest transitions. The order edges correspond to relative displacements of $\pm$0.5. Transitions are colour-coded depending on their wavelength. No correlation or significant structure is apparent, and uncertainties limit the possible amplitude of residual intra-order distortions in the combined sub-spectra to $\la$50\,\ms.}\label{pos_vs_vel}
\end{figure}

\Fref{pos_vs_vel} shows the velocity shift measured for each transition versus its position along its echelle order. The typical uncertainties on the velocity shifts are $\approx$60\,\ms, below the expected level for intra-order distortions (in individual exposures). However, there is no evidence for additional scatter in the velocity shifts beyond that expected from the individual error bars. That is, we do not find evidence in \Fref{pos_vs_vel} for intra-order distortions in the combined spectrum. Therefore, our estimate of the systematic error from intra-order distortions in \Sref{intra-order_sys}, $\approx$0.19\,ppm (see \Tref{systematics}), may well be conservative. \Fref{pos_vs_vel} also demonstrates that the cumulative effect of all systematic effects that may spuriously shift the transitions with respect to each other does not exceed $\approx$60\,\ms\ and, therefore, will not have a significant impact on our measurement of $\varal$ in this system.

\subsubsection{Isotopic abundance variations}\label{without_Mg}

To measure the correct velocities of individual components we must model transitions using their known isotopic structures and correct relative isotopic abundances. However, the isotopic abundances in the absorbers are unknown and we assume the terrestrial abundance pattern for each species \citep{2014MNRAS.438..388M}. If the relative isotopic abundances are substantially different in the absorbers, this will lead to small, spurious shifts measured between the transitions of interest and, therefore, systematic errors in $\varal$ measurements. Some transitions, particularly of light species, and especially \ion{Mg}{ii}, have widely separated ($\sim$1\,\kms) isotopic components, while most others, such as transitions of \ion{Fe}{ii}, have negligible separation between the isotopic components. Therefore, the possibility of differing \ion{Mg}{ii} isotopic abundances (compared to terrestrial) is potentially an important systematic effect for $\varal$ measurements. This effect was explained and quantified in detail in \citet{2001MNRAS.327.1223M} and \citet{2005MNRAS.358..468F}.
 
In a typical absorption system, in which the 3 strongest \ion{Mg}{i/ii} transitions and several of the 5 \ion{Fe}{ii} transitions at 2300--2600\,\AA\ are used to measure $\varal$, \citet{2004LNP...648..131M} showed that if the heavy isotope fraction of Mg ($^{25+26}$Mg/$^{24}$Mg) is zero, instead of 0.21 (the terrestrial value), we should measure $\varal$ to be 4\,ppm too low. If the heavy isotope fraction is 1, we should measure $\varal$ to be too high by 17.5\,ppm. Such variations in isotopic abundances are not implausible \citep{2001MNRAS.327.1223M} so it is important to gauge their possible effect on our measured $\varal$ value.

We checked this possibility by removing the influence the Mg transitions have on $\varal$ in our analysis of the right region. The Mg transitions were kept in the fit so that they still helped to constrain the velocity structure. However, they were effectively decoupled from influencing the $\varal$ parameter by introducing a freely-fit velocity shift between each Mg transition and all other transitions in the fit. The $\chisq$ minimization procedure was started from our fiducial fit in the right region but with $\varal$ reset to 0\,ppm. The resulting best-fit $\varal$ value was $-0.68\pm0.60_{\rm stat}$\,ppm, which is very close to our fiducial value, $\varal=-0.92\pm0.59_{\rm stat}$\,ppm. Indeed, this comparison demonstrates that the Mg transitions have very little weight in constraining $\varal$ in the right region: the statistical uncertainty is almost unaffected when decoupling Mg from $\varal$, primarily because the main spectral features are saturated in \ion{Mg}{ii}. Therefore, we conclude that isotopic abundance variation, at least in Mg, does not influence our result significantly. This also indicates that fringing (see \Sref{fitt}), which is most prominent for the Mg transitions, also has a negligible effect on our $\varal$ measurement.
  
\subsection{Final result for the entire absorption system}\label{comb_err} 

Here we combine the results and error budgets from the three regions to derive our final result for the entire absorption system. The effect of the long-range distortions on $\varal$ is correlated across the 3 regions because it influences the wavelength scale similarly at the position of each transition regardless of which region we consider. Furthermore, the intra-order distortions have a correlated influence on $\varal$ because the transitions are much narrower than echelle orders. We take these correlated systematic errors into account using a Monte Carlo simulation approach, similar to that in \citet{2014MNRAS.445..128E}, to calculate the distribution of final $\varal$ values, the width of which reflects the full error budget, including the correlated systematic effects. 

For each realization of the Monte Carlo simulation, we draw a random $\varal$ value, $(\varal)^i_{\rm rand}$, from a Gaussian distribution for each region $i$. For a given region, the Gaussian is centred on its final $\varal$ value (\Tref{results}) and its 1$\sigma$ width is the quadrature sum of the statistical (\Tref{results}) and uncorrelated systematic error terms (\Tref{systematics}; specifically, the redispersion, velocity structure and convergence uncertainties). To incorporate the correlated effects of uncertainties in our long-range distortion corrections, we draw a random value from a single Gaussian distribution, centred at zero and width $\sigma=1$, and scale it by the long-range distortion uncertainty in each region (\Tref{systematics}). This produces 3 correlated values of $(\varal)^i_{\rm long}$, one for each region $i$. Similarly, for the intra-order distortions, we derive a value of $(\varal)^i_{\rm intra}$ for each region, $i$, which is correlated with that for the other regions. The simulated value of $\varal$ in each region is then $(\varal)^i_{\rm sim} = (\varal)^i_{\rm rand}+ (\varal)^i_{\rm long} + (\varal)^i_{\rm intra}$. The simulated value for the entire absorption system is then the weighted sum of $(\varal)^i_{\rm sim}$ across the 3 regions, with the weight in each region equal to the inverse sum of variances from all error sources in that region (Tables 4 \& 5).

The procedure above is repeated for a large number of realizations to form a (symmetric), Gaussian-like distribution of weighted mean $\varal$ values. This distribution has a mean of $\varal=-1.42$\,ppm and a width $\sigma_{\rm tot} = 0.85$\,ppm which represents the total uncertainty, including statistical and systematic error components. The statistical component of this is just the normal error in a weighted mean of the 3 regions considering only the statistical errors, i.e.~$\sigma_{\rm stat}=1/\sqrt{\sum_i1/(\sigma^i_{\rm stat})^2}=0.55$\,ppm. The systematic error component is therefore the quadrature difference, $\sigma_{\rm sys}=\sqrt{\sigma^2_{\rm tot}-\sigma^2_{\rm stat}}=0.65$\,ppm. This gives the final result for the entire absorption system (also provided in \Tref{results}),
\begin{equation}\label{eq:final}
\varal = -1.42 \pm 0.55_{\rm stat} \pm 0.65_{\rm sys}\,.
\end{equation}
This final result shows no strong evidence for a deviation from the current, laboratory value of $\alpha$ in the absorption cloud. The result is consistent with zero, at the 1.7$\sigma$ level, with a total uncertainty of 0.85\,ppm.       

In \Tref{systematics} we also present the contribution that each type of systematic error makes to the $\sigma_{\rm sys}$ for the whole absorption system. For each type of systematic error, $s$, the same Monte Carlo approach above is followed but with that type of error removed from the process. This provides a new, smaller value of the total error, $\sigma^{\cancel{s}}_{\rm tot}$. The contribution for each type of systematic error is calculated as the quadrature difference $\sigma^s_{\rm sys} = \sqrt{\sigma^2_{\rm tot}-(\sigma^{\cancel{s}}_{\rm tot})^2}$. For example, when removing the long-range distortions from the Monte Carlo simulation, we found the simulated $\varal$ distribution had a width of 0.61\,ppm. The quadrature difference between the total error budget (0.85\,ppm) and this reduced value is 0.59\,ppm. This contribution from uncertainties in the long-range distortion corrections is the dominant systematic error term. The next largest contribution (0.19\,ppm) is from intra-order distortions. This demonstrates the importance of removing these sources of systematic error for future, more precise measurements of $\varal$ using new spectrographs and telescopes (see \Sref{artifacts} for discussion).

\section{Discussion}\label{discussion}

\begin{table*}
\caption{ Comparison of the constraints on $\varal$ from the $\zab=1.1508$ absorption system towards QSO HE 0515$-$4414 in the literature \citep[][]{2004A&A...415L...7Q, 2006A&A...449..879L, 2006A&A...451...45C, 2008EPJST.163..173M} with our measurements. The independence and overlap between different data sets used is defined in the column `Exposures used'. The $\SN$ ratio reported corresponds to the continuum near the \ion{Fe}{ii}\,2600 transition and is expressed on a per pixel basis. L06 and M08b report $\SN$ ratio for the individual exposures that they use in their analysis. The value reported in this table is the quadrature sum of these individual values. Pixel sizes for the UVES and HARPS spectra used in this work are 1.3 \kms\ and 0.85 \kms, respectively. The column labeled `Components' represents the number of velocity components fitted in the right region; in our analysis we fitted 32 and 25 components to the left and central regions, respectively.}\label{previous}
\centering
  \begin{tabular}{@{}l c c c c c c c}   
  \hline
Reference  &   Instrument   & Exposure    &Exposures             & $\SN$              & Resolving             & Components   & $\varal$ [ppm]                                      \\
           &                & time [s]    &used$^a$              &                    & power                 &              &                                                      \\
\hline                                                                                                                                                                                           
Q04        &   UVES           & $53100$   & 1      & $\sim$130          & $55000$               & 12           & $-0.40 \pm 1.90_{\rm stat} \pm 2.70_{\rm sys}$       \\
L06        &   UVES           & $32400$   & 2      & $\sim$135          & $55000$               & 13           & $-0.07 \pm 0.84_{\rm stat}$                            \\
M08b       &   UVES           & $32400$   & 2      & $\sim$135          & $55000$               & 13           & $-0.12 \pm 1.79_{\rm stat}$                             \\
C06        &   HARPS          & $78600$   &                      & $\sim$35           & $112000$              & 15           & $+0.50 \pm 2.40_{\rm stat}$                            \\
C06        &   UVES           & $53100$   & 1      & $\sim$130          & $55000$               &  9           & $+1.00 \pm 2.20_{\rm stat}$                            \\
This work  &   UVES           & $155800$  & 3     & $\sim$250          & $53500$--$93300$      & 49           & $-1.42\pm0.55_{\rm stat}\pm0.65_{\rm sys}$                      \\
  \hline
\multicolumn{8}{l}{$^a$Sub-spectra:} \\
\multicolumn{8}{l}{1: All exposures from our 2000 sub-spectrum;} \\
\multicolumn{8}{l}{2: Three exposures from 1999 sub-spectrum, which we have not used in this work because they do not have attached ThAr exposures,} \\ 
\multicolumn{8}{l}{and 4 out of 13 exposures from our 2000 sub-spectrum;} \\
\multicolumn{8}{l}{3: All five sub-spectra (see details in \Sref{UVES_ORC}).} \\
  \end{tabular}
\end{table*}

\subsection{Comparison with previous measurements of $\varal$ in the same absorption system}\label{comp_pre_mea}

Comparison of our result with previous measurements of $\varal$ in the same absorption system, reported in Q04, L06, C06 and M08b, is presented in \Tref{previous}. All of these previous measurements are consistent with our new measurement and with zero variation. However, in \Sref{DC_HARPS} we used the HARPS spectrum to demonstrate that the spectra used in the previous studies were all affected by the long-range distortions of 110--220\,\ms\ per 1000\,\AA. These studies used the comparison of \ion{Fe}{ii}\,1608 with the redder \ion{Fe}{ii} transitions to constrain $\varal$ so the long-range distortions should have caused systematic shifts in $\varal$ of order $\approx$2\,ppm, as we found for our constraint in \Sref{long-range_sys}. That is, if corrected, these previous measurements would likely be larger by $\approx$2\,ppm. This shift is generally similar to the uncertainties in the values of $\varal$ derived from UVES spectra by Q04, L06, C06 and M08b. In this work we corrected the wavelength scales of the previous and new UVES spectra for these distortions, leaving a residual contribution to the systematic error budget of only 0.6\,ppm (see \Tref{systematics}).

\Tref{previous} also shows how our new fit contains a much larger number of velocity components than previous fits. Our fit to the entire absorption system has 106 velocity components, but the only region common to all previous studies corresponds to our right region, in which \Tref{previous} shows we fitted 49 components. This is more than 3 times larger than in any previous fit in that region. We demonstrated in \Sref{fitt} that our fitting approach was not able to strictly satisfy the normal information criterion of minimizing $\chisqn$. In this sense, our 106 fitted velocity components may represent a lower limit to the actual number in the absorption profile. Indeed, we stopped adding components to avoid compensating for data artifacts we identified (e.g.~fringing). Our analysis of the higher-resolution HARPS and bHROS spectra in \Sref{unresolved} also provides no evidence that the absorption profile contains fewer components than we fit. It therefore appears that previous analyses significantly underestimated the number of components, even taking into account the lower $\SN$ of the spectra used in those studies \citep[see also discussion of Q04's fit in][]{2008MNRAS.384.1053M}. \citet{2008MNRAS.384.1053M,2008Sci...320.1611M} and \citet{2015MNRAS.454.3082W} have shown that under-fitting in this way often leads to spurious shifts in varying-constant measurements from absorption profiles. On the other hand, \citet{2008MNRAS.384.1053M} showed that over-fitting, to the very small extent allowed by $\chisq$ minimization codes like {\sc vpfit}, yields accurate $\varal$ values with slightly overestimated statistical errors.

The statistical errors reported by Q04 and C06 are significantly larger than ours, as expected considering that the total $\SN$ of our spectra is $\approx$2 times that in those studies. However, the statistical error quoted by L06 is similar to ours. \citet{2008MNRAS.384.1053M} demonstrated that the ``limiting precision" on $\varal$ available from the spectra used by L06 was worse than the statistical uncertainty quoted in that study. M08b revised the L06 measurement by taking into account the correlations between the series of pair-wise velocity measurements between transitions used to measure $\varal$. They reported an increased statistical error of 1.79\,ppm, which is in agreement with the limiting precision calculated by \citet{2008MNRAS.384.1053M}. Following the formalism in \citet{2008MNRAS.384.1053M}, the limiting precision on $\varal$ available from the $\SN$ of our UVES spectra in the right region is 0.34\,ppm, which is significantly smaller than our measured statistical error, 0.59\,ppm, as expected.

Finally, as detailed in \citet{2008MNRAS.384.1053M}, the measurement of $\varal$ from both the UVES and HARPS spectra studied in C06 was compromised by problems in the $\chisq$ minimization process. On the other hand, the algorithm used in {\sc vpfit} has been tested extensively with Monte Carlo Markov Chain analyses \citep{2009MmSAI..80..864K} and shown to return accurate (even slightly conservative) statistical errors on $\varal$.

\subsection{Comparison of our results with other previous $\varal$ measurements}

\begin{figure}
  \centering
    \includegraphics[width=0.49\textwidth,natwidth=610,natheight=642]{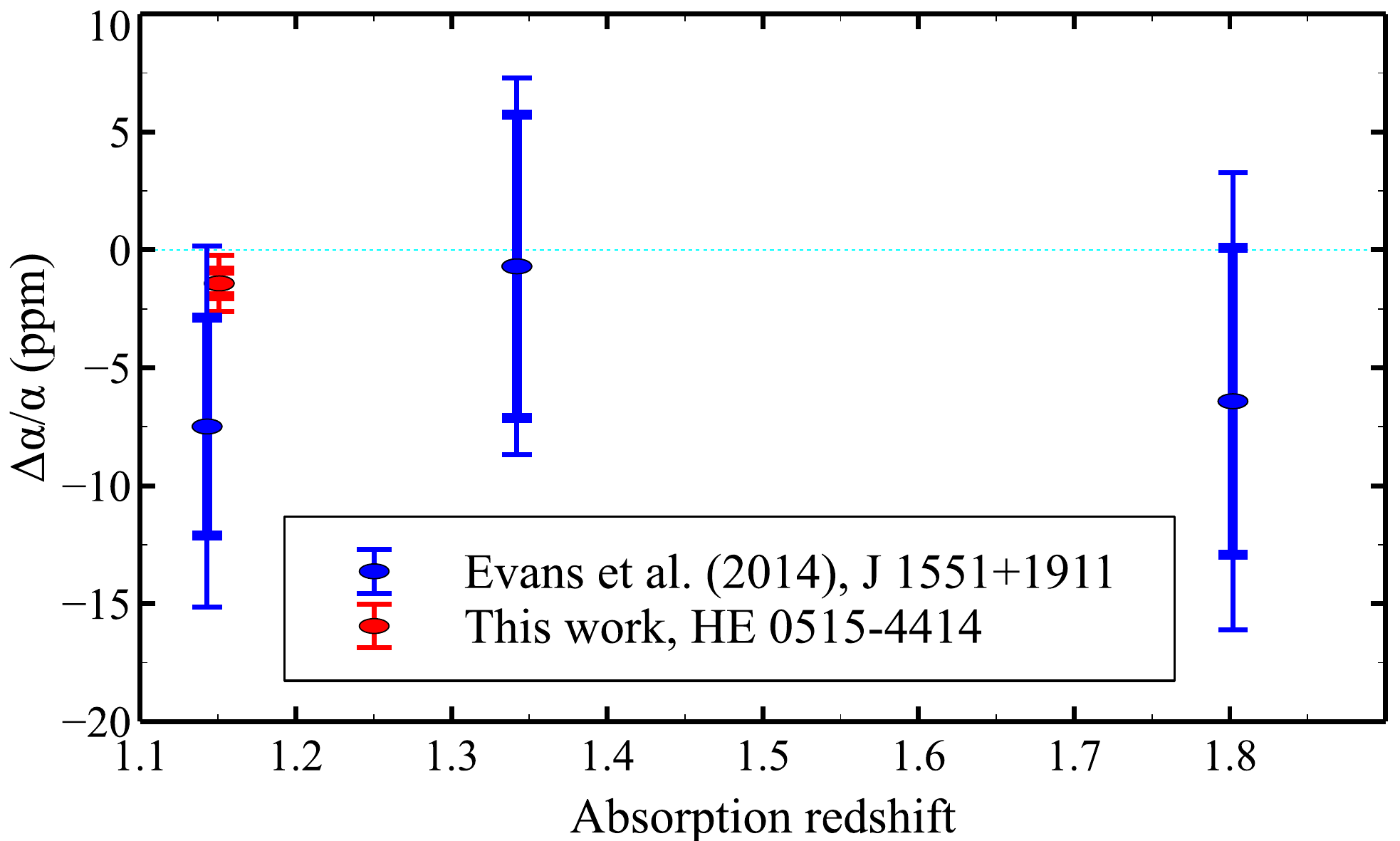}
  \caption{Comparison of our measurement with the combined measurements for each absorber from \citet{2014MNRAS.445..128E}. Thick error bars represent statistical errors and thin error bars represent systematic errors (that is, the error from quadrature combination of these error components is smaller than the total error bar represented on this figure).}\label{comparison}
\end{figure}

The only other $\varal$ measurements from absorption lines that have been corrected for the long-range distortions are from a single quasar sight-line with 3 absorption systems, studied using the Keck, VLT and Subaru telescopes by \citet{2014MNRAS.445..128E}. \Fref{comparison} compares our new measurement with the best estimate of $\varal$ in each of those 3 absorbers (averaged across the 3 independent spectra). Our new result is clearly consistent with all three previous measurements. Our combined result of $-1.42\pm0.55_{\rm stat}\pm0.65_{\rm sys}$ differs by $1.1\sigma_{\rm comb}$ from the combined \citet{2014MNRAS.445..128E} result of $-5.40\pm3.25_{\rm stat}\pm1.53_{\rm sys}$, where $\sigma_{\rm comb}$ is the quadrature sum of the statistical and systematic errors in both studies. The statistical and systematic errors are 5.9 and 2.4 times smaller, respectively, in our new measurement compared with the combined measurement from \citet{2014MNRAS.445..128E}.

Comparison with other previous measurements of $\varal$ from quasar spectra is very difficult because they will all have been affected by long-range wavelength distortions; none of them is corrected for this important systematic error and, therefore, none can be considered reliable. This includes the large statistical samples from the VLT \citep{2012MNRAS.422.3370K} and Keck \citep{2004LNP...648..131M} which showed significant deviations in $\varal$ from zero. Indeed, \citet{2015MNRAS.447..446W} used archival supercalibration exposures over two decades to demonstrate that a simple model of the long-range distortions can adequately explain the VLT results and partially explain those from Keck.

Nevertheless, disregarding those important concerns, the uncorrected VLT and Keck results together imply a dipole-like variation of $\alpha$ across the sky \citep{2012MNRAS.422.3370K} which can, in principle, be ruled out with new, distortion-corrected measurements like ours. The simplest model proposed by \citet{2012MNRAS.422.3370K} is a dipole (with no monopole term) in which $\varal=10.2_{-1.9}^{+2.2}\cos(\Theta)$\,ppm where $\Theta$ is the angle between the dipole direction (${\rm RA}=17.4\pm0.9$\,h, ${\rm Dec.}=-58\pm9^\circ$) and a selected position on the sky. For the direction towards HE 0515$-$4414, $\Theta=77.8^\circ$ and the model implies $\varal$ should be $2.2\pm1.6$\,ppm. Our measured value differs by $2.0\sigma_{\rm comb}$ from this expectation. If a monopole term is allowed in the dipole model, \citet{2012MNRAS.422.3370K} find it to be $-1.8$\,ppm -- similar to our total uncertainty -- and the expected value in the direction of HE 0515$-$4414 becomes $0.8\pm1.9$\,ppm, differing by $1.1\sigma_{\rm comb}$ from our measurement. That is, our new measurement cannot be used to confirm or rule out the dipole explanation for the non-zero Keck and VLT large-sample results, mainly because the direction to HE 0515$-$4414 is not well aligned with the dipole (or anti-pole) direction. However, a small number of very precise measurements like ours, towards quasars located $\Theta\approx90^\circ$ away from that direction would certainly have that potential.

\subsection{Next-generation spectrographs and telescopes: expected systematic effects}\label{artifacts}
Our measurement of $\varal$ includes four different systematic uncertainty terms. We have also identified different data artifacts that can cause systematic shifts in $\varal$ measurements. To prepare for the high quality of spectra expected from the next generation of telescopes and spectrographs, it is important to assess all of these effects in that context. This will help us understand which of these effects will still be present in future measurements and what improvements can be made to the spectrographs, telescopes and data reduction pipelines, so that a better measurement of $\varal$ can be made.

One category of effects is associated with CCDs and data reduction techniques: fringing effects and artifacts from flux extraction, sky subtraction and flat-fielding. Avoiding or mitigating these effects would require using new types of detectors which are not prone to these problems and/or new approaches to reducing echelle spectra. These problems may be addressed on the longer timescales involved in preparing for and building G-Clef on the Giant Magellan Telescope (GMT) \citep{2014SPIE.9147E..26S} and HIRES on the European Extremely Large Telescope (E-ELT) \citep{2014SPIE.9147E..23Z}. These telescopes will have much higher light collecting area which will enable significant improvements in $\SN$ ratio and, therefore, statistical uncertainties in $\varal$ measurements. Therefore, it is very important to reduce systematic effects and data artifacts commensurately.

In the short-term future, the next spectrograph that will improve $\varal$ measurements will be the Echelle SPectrograph for Rocky Exoplanet and Stable Spectroscopic Observations (ESPRESSO), to be mounted on the VLT \citep{2010SPIE.7735E..0FP}. ESPRESSO will be thermally and pressure stabilized and will use optical fibres instead of a slit, similar to the HARPS spectrograph. Therefore, spectra taken with ESPRESSO should be free from long-range wavelength distortions. Furthermore, it will be calibrated with laser frequency combs \citep{2007MNRAS.380..839M, 2010SPIE.7735E..0FP}, so we do not expect it to be prone to intra-order distortions either \citep{2010MNRAS.405L..16W}. In this work, the systematic error terms associated with the uncertainties in the long-range wavelength distortion corrections and the intra-order wavelength distortions contribute the most to our systematic error budget. If we exclude these two terms, which we could do if we were using ESPRESSO spectra instead of UVES, our systematic error would be reduced by a factor of three to $\sim$0.2\,ppm.

Using ESPRESSO to reobserve quasar spectra that have already been observed with UVES to avoid long-range and intra-order distortions will require approximately the same observing time as already invested. For example, to achieve the same statistical precision from the $\zab=1.1508$ absorption system towards HE 0515$-$4414 as we achieve here would require approximately 4 nights of ESPRESSO observations. To be more specific, due to ESPRESSO's wavelength coverage (3800--6860\,\AA), which does not cover \ion{Fe}{ii}\,1608, the statistical precision on $\varal$ measurements will weaken to $\approx$1.8\,ppm so substantially more observing time would be needed to compensate for this. This means that quasars with appropriate absorption systems that fall between $\zab\sim1.37$ and $\sim$1.85 should be chosen for observations with ESPRESSO. Similarly, the quasar providing the $\varal$ measurement with the next smallest statistical error (2.4\,ppm), HE 2217$-$2818 \citep{2013A&A...555A..68M}, would require 20 hours of new ESPRESSO observations. The absorption system in the sight-line towards this quasar falls into the aforementioned redshift range so the new observations do not need to be much longer than those of \citet{2013A&A...555A..68M}.

An alternative approach would be to recalibrate the existing UVES spectra with new ESPRESSO observations and the DC method comparison, just as we used the HARPS spectra to correct the long-range distortions in the UVES spectra of HE 0515$-$4414. In this approach, new ESPRESSO spectra would only require a relatively low $\SN$ ratio, and consequently shorter total exposure times, because they would be used primarily to correct the long-range distortions in the UVES spectra. The key assumptions in this approach are that the long-range distortions of UVES spectra are approximately linear and with similar slopes in the blue and red arms, which was shown to be typical by \citet{2015MNRAS.447..446W}, and that the intra-order distortions do not contribute a large systematic error (as demonstrated in this paper). Provided these assumptions, an ESPRESSO spectrum with a $\SN$ of $\sim$35\,per\,\kms, similar to the HARPS spectrum in this paper, would allow the long-range distortions in the UVES spectra to be corrected and contribute a residual long-range distortion systematic error of only $\sim$0.6\,ppm to $\varal$. This could be achieved with just 2--3 hours of ESPRESSO observations for bright objects like HE 0515$-$4414.

The main advantage of this approach is in cases where the existing UVES spectra provide a statistical uncertainty on $\varal$ of $\ga$1\,ppm -- in these cases, a residual calibration error in $\varal$ of $\sim$0.6\,ppm would not be dominant (though the assumptions above must be carefully assessed). This includes all absorbers previously observed except for the one studied here towards HE 0515$-$4414. For example, by correcting the previously studied UVES spectra of HE 2217$-$2818 \citep{2013A&A...555A..68M} with just $\sim$3 hours of ESPRESSO observations, a residual calibration error of 0.6\,ppm would be very small compared to the statistical error of 2.4\,ppm from the UVES spectra of the $\zab=1.6919$ absorber. The assumption that the blue and red arm long-range distortion slopes are the same is not required for that absorber because the \ion{Fe}{ii}\,1608 transition falls in the red UVES arm.

Approximately 20 other quasars with $\SN>100$\,per pixel are available from the UVES data archive mainly from two previous Large Programs. If we adopt the method proposed in the previous paragraph, we could recalibrate the existing spectra of all 20 quasars by observing for only $\sim$40\,h with ESPRESSO. For comparison, 40 hours of exposures would enable a $\SN$ of $\sim$100 for just two quasars with entirely new ESPRESSO observations. That is, using the existing UVES spectra of these quasars, we could make the first sample of reliable measurements very efficiently, with this alternative approach. Most of the absorbers in these UVES spectra already have modeled velocity structures, which will make the analysis even simpler.

\subsection{Improving $\SN$ ratio or resolution?}

From the perspective of designing future spectrographs for measuring $\varal$ it is crucial to understand whether it is more important to improve resolution or $\SN$\,per \kms. The key for obtaining a \emph{reliable} $\varal$ measurement is the ability to decompose the metal absorption profiles accurately; once that is achieved, the statistical precision is determined by the $\SN$\,per \kms\ of the spectrum. It is often implicitly assumed, and sometimes stated (e.g.\ C06), that absorption profiles are more easily decomposed into sub-components, and/or the precision on $\varal$ is improved, if the resolution is improved. However, as discussed below, the extent to which that is true depends critically on the $\SN$ ratio and to some extent on the nature of the velocity structure itself.

In \Sref{unresolved} we addressed this issue by examining three different spectra with different resolutions and $\SN$\,per \kms. We were not able to identify additional velocity structure in the higher-resolution spectra (from HARPS and bHROS), mainly because of their very low $\SN$\,per \kms, particularly for bHROS. This effectively hid any information about additional velocity structure. The very firm conclusion from this comparison is that at least similar $\SN$\,per \kms\ is necessary in the higher-resolution spectra of this absorber to enable identification of any additional velocity structure. Furthermore, it is clear that obtaining, for example, a HARPS spectrum with the same $\SN$\,per \kms\ as our UVES spectrum of HE0515$-$4414 would not provide a substantially improved precision on $\varal$. 

The main reason for the conclusion above was that the velocity components are packed closely together, relative to their $b$-parameters, in the particular absorber studied here. This implies that increasing the resolving power does not assist in decomposing the absorption profile; any additional structure revealed will be subtle and, therefore, will not provide a substantially more precise $\varal$ measurement. There are also possibly \emph{harmful} effects of moving to higher resolution. For example, consider a spectrograph with adjustable resolving power, from $R=60000$ (typical of UVES) to $R=120000$ (like bHROS and the lower-resolution mode of ESPRESSO), or even to $R=240000$ (the high-resolution mode of ESPRESSO). For absorption systems like the one studied here, in which the velocity components are evidently closely packed, moving from $R=60000$ to 120000 would not provide much increase in precision on $\varal$. However, if the spectrograph is fibre-fed then achieving the higher resolution mode may involve spreading the quasar light across a larger number of spatial pixels. This was the case in bHROS (see \Sref{bHROS_ORC}), where the spatial extent of the quasar signal was $\approx$50 pixels. This increases the read-noise per spectral pixel so that observing for the same time at $R=60000$ and $R=120000$ provides a lower $\SN$\,per \kms\ at $R=120000$. This ultimately decreases the precision on $\varal$, a loss that may well outweigh the small gain from the increased resolving power. The loss may be even greater for the $R=240000$ mode. These conclusions will only apply to other absorption systems if their velocity components are distributed similarly to those of HE0515$-$4414, i.e.\ packed closely together relative to their $b$-parameters.

\subsection{Measuring $\varal$ from a single absorption system versus an ensemble of absorption systems}\label{1vsmany}

Our measurement of $\varal$ is currently the most precise from any absorption system towards any quasar, and is comparable to the ensemble precision from the large samples of Keck and VLT spectra from \citet{2004LNP...648..131M} and \citet{2012MNRAS.422.3370K} ($\sim$1.1\,ppm). However, the fact that this measurement is made in a single absorption system has a potentially important disadvantage: many systematic effects can be expected to be random in sign and magnitude from absorber to absorber, so analysing a single system requires careful treatment of the systematic errors presented here. For example, apart from the possibility of isotopic abundance variations, all of the systematic errors discussed here (see \Tref{systematics}) would randomize in a large sample of absorbers. In principle, it is also possible that we have missed a potential source of systematic errors, even though we have empirically corrected the wavelength scale using the HARPS spectra, and the consistency checks in \Sref{consistency_checks} did not reveal any significant, unknown effects. Therefore, a larger sample of measurements, with similar precision to that obtained here, would be desirable.

However, at present the UVES spectrum of HE 0515$-$4414 has much higher $\SN$ than any other in the UVES archive. Very few ($\la$10) archival UVES spectra have $\SN>120$\,per pixel, but another $\sim$10 have $\SN\sim100$. Assuming that $\sim$20 $\varal$ measurements can be made from UVES spectra with peak $\SN\sim100$, a precision of $\sim$0.55\,ppm\,$\times250/100/\sqrt{20} = 0.3$\,ppm would be available (taking our spectrum's peak \SN\ to be 250\,per pixel). This is similar to our measurement's statistical precision (0.55\,ppm) but, if long-range distortions were corrected in each of these $\sim$20 new measurements, the main advantage would be that the remaining systematic errors would randomize to a negligible residual value. Significantly reducing the statistical error budget would require a much larger number of $\SN\sim100$ spectra, and obtaining a sample of $\sim$20 much higher $\SN$ spectra, like that of HE 0515$-$4414 used here, does not seem possible in the next decade.

\section{Conclusions}

In this work we have measured the relative variation in the fine-structure constant in the $\zab=1.1508$ absorber towards the quasar HE 0515$-$4414. Because of its brightness ($V\approx14.9$\,mag) this quasar has been observed frequently with UVES, so a large number of archival spectra are available ($\approx$30\,h). Here we add 13\,h of new UVES exposures to obtain a total $\SN\approx250$\,per 1.3-\kms\ pixel (at its peak), the highest for any echelle spectrum of a quasar at $\zem>1$. This, and the large number of narrow features in the \ion{Mg}{i/ii} and \ion{Fe}{ii} absorption profiles, provide a very small statistical uncertainty on $\varal$ of 0.55\,parts per million (ppm).

Most importantly, we have corrected a large systematic error in the UVES spectra, from long-range distortions of the wavelength calibration, by using a HARPS spectrum of the same quasar. Left uncorrected, these would have caused a spurious shift in $\varal$ of $\approx$2.1\,ppm. However, by directly comparing the UVES and HARPS spectra the correction leaves a residual systematic uncertainty of just 0.59\,ppm. This assumes that the distortions are linear with wavelength and that they have the same slope in the red and blue arms of the UVES spectrograph. Previous studies of these distortions in UVES spectra generally support these assumptions \citep{2015MNRAS.447..446W}. Other systematic errors, mainly from short-range (i.e.\ intra-order) distortions and uncertainties in the absorber's velocity structure, contribute a further 0.26\,ppm uncertainty. A series of consistency checks suggest that our total systematic error budget of 0.65\,ppm is reliable and that astrophysical systematic errors, such as isotopic abundance variations, are unimportant.

Our final result for this absorber is $\varal=-1.42\pm0.55_{\rm stat}\pm0.65_{\rm sys}$\,ppm. This is consistent with no variation in the fine-structure constant and is the most precise measurement from a single absorption system to date. The precision is comparable to the ensemble precision from the large Keck and VLT samples of absorption systems studied by \citet{2004LNP...648..131M} and \citet{2012MNRAS.422.3370K}. It is unlikely that measurements of $\varal$ in other, individual absorption systems, will match the precision obtained here until new 25--40-m telescopes become available. Indeed, this work provides a preview of effects that must be addressed in the very high $\SN$ spectra from those future telescopes. For example, accurate knowledge of the resolving power was required to model the absorption profile in this work, and CCD and/or data-reduction artifacts were evident in our $\SN\sim250$\,pix$^{-1}$ spectrum.

Finally, given that all previous $\varal$ measurements \citep[except those of][]{2014MNRAS.445..128E} will have been significantly affected by long-range distortions, it is crucial to obtain new measurements which are corrected for (or resistant to) this important systematic effect. In this context, the upcoming ESPRESSO spectrograph on the VLT is very important because it should provide spectra free of this effect (and the intra-order distortions). This work provides two insights for using ESPRESSO most efficiently for precise and reliable $\varal$ measurements. Firstly, the recalibration approach demonstrated in this work could be applied: existing, high-$\SN$ UVES spectra could be recalibrated with relatively short ESPRESSO observations ($\sim$2--3\,h) of the same quasars. This would produce a sample of $\sim$20 reliable, high-precision measurements in $\sim$20\,per cent of the time required to build the same $\SN$ with all new ESPRESSO spectra. Secondly, comparison of the UVES and higher-resolution HARPS and bHROS spectra of HE 0515$-$4414 implies that the absorber comprises many closely-packed, narrow velocity components; the increased resolution provides little additional information about the velocity structure or a significant increase in precision on $\varal$. Therefore, if the velocity structures of most other absorbers also comprise many closely-packed, narrow components, it is unlikely that a resolving power $R>100000$ will benefit varying-$\alpha$ measurements, especially if $\SN$ is compromised to obtain the higher resolution.

\section*{Acknowledgments}
We would like to thank Tyler Evans, Adrian Malec and Neil Crighton for advice and support in various stages of this work and use of their codes. SMK has been supported in part by Swinburne University of Technology Postgraduate Research Award. MTM thanks the Australian Research Council for Discovery Projects grant DP110100866 which supported this work. This work is based mainly on observations carried out at the European Southern Observatory under programs No. 60.A-9022(A), 66.A-0212(A), 072.A-0100(A), 079.A-0404(A) and 082.A-0078(A), with the UVES spectrograph installed at the Kueyen UT2 on Cerro Paranal, Chile.

\vspace{-1.5em}
\bibliographystyle{mn2e}
\bibliography{paper_arXiv}

\begin{thebibliography}{}
\makeatletter
\relax
\def\mn@urlcharsother{\let\do\@makeother \do\$\do\&\do\#\do\^\do\_\do\%\do\~}
\def\mn@doi{\begingroup\mn@urlcharsother \@ifnextchar [ {\mn@doi@}
  {\mn@doi@[]}}
\def\mn@doi@[#1]#2{\def\@tempa{#1}\ifx\@tempa\@empty \href
  {http://dx.doi.org/#2} {doi:#2}\else \href {http://dx.doi.org/#2} {#1}\fi
  \endgroup}
\def\mn@eprint#1#2{\mn@eprint@#1:#2::\@nil}
\def\mn@eprint@arXiv#1{\href {http://arxiv.org/abs/#1} {{\tt arXiv:#1}}}
\def\mn@eprint@dblp#1{\href {http://dblp.uni-trier.de/rec/bibtex/#1.xml}
  {dblp:#1}}
\def\mn@eprint@#1:#2:#3:#4\@nil{\def\@tempa {#1}\def\@tempb {#2}\def\@tempc
  {#3}\ifx \@tempc \@empty \let \@tempc \@tempb \let \@tempb \@tempa \fi \ifx
  \@tempb \@empty \def\@tempb {arXiv}\fi \@ifundefined
  {mn@eprint@\@tempb}{\@tempb:\@tempc}{\expandafter \expandafter \csname
  mn@eprint@\@tempb\endcsname \expandafter{\@tempc}}}

\bibitem[\protect\citeauthoryear{{Agafonova}, {Molaro}, {Levshakov}  \&
  {Hou}}{{Agafonova} et~al.}{2011}]{2011A&A...529A..28A}
{Agafonova} I.~I.,  {Molaro} P.,  {Levshakov} S.~A.,   {Hou} J.~L.,  2011,
  \mn@doi [\aap] {10.1051/0004-6361/201016194}, \href
  {http://adsabs.harvard.edu/abs/2011A%26A...529A..28A} {529, A28}

\bibitem[\protect\citeauthoryear{{Bagdonaite}, {Ubachs}, {Murphy}  \&
  {Whitmore}}{{Bagdonaite} et~al.}{2014}]{2014ApJ...782...10B}
{Bagdonaite} J.,  {Ubachs} W.,  {Murphy} M.~T.,   {Whitmore} J.~B.,  2014,
  \mn@doi [\apj] {10.1088/0004-637X/782/1/10}, \href
  {http://adsabs.harvard.edu/abs/2014ApJ...782...10B} {782, 10}

\bibitem[\protect\citeauthoryear{{Carswell} \& {Webb}}{{Carswell} \&
  {Webb}}{2014}]{2014ascl.soft08015C}
{Carswell} R.~F.,  {Webb} J.~K.,  2014, {VPFIT: Voigt profile fitting program},
  Astrophysics Source Code Library (\mn@eprint {ascl} {1408.015})

\bibitem[\protect\citeauthoryear{{Chand}, {Srianand}, {Petitjean}  \&
  {Aracil}}{{Chand} et~al.}{2004}]{2004A&A...417..853C}
{Chand} H.,  {Srianand} R.,  {Petitjean} P.,   {Aracil} B.,  2004, \mn@doi
  [\aap] {10.1051/0004-6361:20035701}, \href
  {http://adsabs.harvard.edu/abs/2004A%26A...417..853C} {417, 853}

\bibitem[\protect\citeauthoryear{{Chand}, {Srianand}, {Petitjean}, {Aracil},
  {Quast}  \& {Reimers}}{{Chand} et~al.}{2006}]{2006A&A...451...45C}
{Chand} H.,  {Srianand} R.,  {Petitjean} P.,  {Aracil} B.,  {Quast} R.,
  {Reimers} D.,  2006, \mn@doi [\aap] {10.1051/0004-6361:20054584}, \href
  {http://adsabs.harvard.edu/abs/2006A%26A...451...45C} {451, 45}

\bibitem[\protect\citeauthoryear{{Dekker}, {D'Odorico}, {Kaufer}, {Delabre}  \&
  {Kotzlowski}}{{Dekker} et~al.}{2000}]{2000SPIE.4008..534D}
{Dekker} H.,  {D'Odorico} S.,  {Kaufer} A.,  {Delabre} B.,   {Kotzlowski} H.,
  2000, in {Iye} M.,  {Moorwood} A.~F.,  eds,  Society of Photo-Optical
  Instrumentation Engineers (SPIE) Conference Series Vol. 4008, Optical and IR
  Telescope Instrumentation and Detectors. pp 534--545

\bibitem[\protect\citeauthoryear{{Dzuba}, {Flambaum}  \& {Webb}}{{Dzuba}
  et~al.}{1999}]{1999PhRvL..82..888D}
{Dzuba} V.~A.,  {Flambaum} V.~V.,   {Webb} J.~K.,  1999, \mn@doi [Physical
  Review Letters] {10.1103/PhysRevLett.82.888}, \href
  {http://adsabs.harvard.edu/abs/1999PhRvL..82..888D} {82, 888}

\bibitem[\protect\citeauthoryear{{Evans} \& {Murphy}}{{Evans} \&
  {Murphy}}{2013}]{2013ApJ...778..173E}
{Evans} T.~M.,  {Murphy} M.~T.,  2013, \mn@doi [\apj]
  {10.1088/0004-637X/778/2/173}, \href
  {http://adsabs.harvard.edu/abs/2013ApJ...778..173E} {778, 173}

\bibitem[\protect\citeauthoryear{{Evans} et~al.,}{{Evans}
  et~al.}{2014}]{2014MNRAS.445..128E}
{Evans} T.~M.,  et~al., 2014, \mn@doi [\mnras] {10.1093/mnras/stu1754}, \href
  {http://adsabs.harvard.edu/abs/2014MNRAS.445..128E} {445, 128}

\bibitem[\protect\citeauthoryear{{Fenner}, {Murphy}  \& {Gibson}}{{Fenner}
  et~al.}{2005}]{2005MNRAS.358..468F}
{Fenner} Y.,  {Murphy} M.~T.,   {Gibson} B.~K.,  2005, \mn@doi [\mnras]
  {10.1111/j.1365-2966.2005.08781.x}, \href
  {http://adsabs.harvard.edu/abs/2005MNRAS.358..468F} {358, 468}

\bibitem[\protect\citeauthoryear{{Griest}, {Whitmore}, {Wolfe}, {Prochaska},
  {Howk}  \& {Marcy}}{{Griest} et~al.}{2010}]{2010ApJ...708..158G}
{Griest} K.,  {Whitmore} J.~B.,  {Wolfe} A.~M.,  {Prochaska} J.~X.,  {Howk}
  J.~C.,   {Marcy} G.~W.,  2010, \mn@doi [\apj] {10.1088/0004-637X/708/1/158},
  \href {http://adsabs.harvard.edu/abs/2010ApJ...708..158G} {708, 158}

\bibitem[\protect\citeauthoryear{{King}, {Mortlock}, {Webb}  \&
  {Murphy}}{{King} et~al.}{2009}]{2009MmSAI..80..864K}
{King} J.~A.,  {Mortlock} D.~J.,  {Webb} J.~K.,   {Murphy} M.~T.,  2009,
  \memsai, \href {http://adsabs.harvard.edu/abs/2009MmSAI..80..864K} {80, 864}

\bibitem[\protect\citeauthoryear{{King}, {Webb}, {Murphy}, {Flambaum},
  {Carswell}, {Bainbridge}, {Wilczynska}  \& {Koch}}{{King}
  et~al.}{2012}]{2012MNRAS.422.3370K}
{King} J.~A.,  {Webb} J.~K.,  {Murphy} M.~T.,  {Flambaum} V.~V.,  {Carswell}
  R.~F.,  {Bainbridge} M.~B.,  {Wilczynska} M.~R.,   {Koch} F.~E.,  2012,
  \mn@doi [\mnras] {10.1111/j.1365-2966.2012.20852.x}, \href
  {http://adsabs.harvard.edu/abs/2012MNRAS.422.3370K} {422, 3370}

\bibitem[\protect\citeauthoryear{{Kotu{\v s}}, {Murphy}  \&
  {Carswell}}{{Kotu{\v s}} et~al.}{2016}]{srdan_kotus_2016_51715}
{Kotu{\v s}} S.~M.,  {Murphy} M.~T.,   {Carswell} R.~F.,  2016, {Quasar spectra
  and absorption profile fits of HE 0515-4414 for limiting fine-structure
  constant variability}, \mn@doi{10.5281/zenodo.51715}, \url
  {http://dx.doi.org/10.5281/zenodo.51715}

\bibitem[\protect\citeauthoryear{{Levshakov}, {Centuri{\'o}n}, {Molaro}  \&
  {D'Odorico}}{{Levshakov} et~al.}{2005}]{2005A&A...434..827L}
{Levshakov} S.~A.,  {Centuri{\'o}n} M.,  {Molaro} P.,   {D'Odorico} S.,  2005,
  \mn@doi [\aap] {10.1051/0004-6361:20041827}, \href
  {http://adsabs.harvard.edu/abs/2005A%26A...434..827L} {434, 827}

\bibitem[\protect\citeauthoryear{{Levshakov}, {Centuri{\'o}n}, {Molaro},
  {D'Odorico}, {Reimers}, {Quast}  \& {Pollmann}}{{Levshakov}
  et~al.}{2006}]{2006A&A...449..879L}
{Levshakov} S.~A.,  {Centuri{\'o}n} M.,  {Molaro} P.,  {D'Odorico} S.,
  {Reimers} D.,  {Quast} R.,   {Pollmann} M.,  2006, \mn@doi [\aap]
  {10.1051/0004-6361:20053827}, \href
  {http://adsabs.harvard.edu/abs/2006A%26A...449..879L} {449, 879}

\bibitem[\protect\citeauthoryear{{Levshakov}, {Molaro}, {Lopez}, {D'Odorico},
  {Centuri{\'o}n}, {Bonifacio}, {Agafonova}  \& {Reimers}}{{Levshakov}
  et~al.}{2007}]{2007A&A...466.1077L}
{Levshakov} S.~A.,  {Molaro} P.,  {Lopez} S.,  {D'Odorico} S.,  {Centuri{\'o}n}
  M.,  {Bonifacio} P.,  {Agafonova} I.~I.,   {Reimers} D.,  2007, \mn@doi
  [\aap] {10.1051/0004-6361:20066064}, \href
  {http://adsabs.harvard.edu/abs/2007A%26A...466.1077L} {466, 1077}

\bibitem[\protect\citeauthoryear{{Malec} et~al.,}{{Malec}
  et~al.}{2010}]{2010MNRAS.403.1541M}
{Malec} A.~L.,  et~al., 2010, \mn@doi [\mnras]
  {10.1111/j.1365-2966.2009.16227.x}, \href
  {http://adsabs.harvard.edu/abs/2010MNRAS.403.1541M} {403, 1541}

\bibitem[\protect\citeauthoryear{{Margheim}}{{Margheim}}{2008}]{2008psa..conf..297M}
{Margheim} S.~J.,  2008, in {Santos} N.~C.,  {Pasquini} L.,  {Correia}
  A.~C.~M.,   {Romaniello} M.,  eds, Precision Spectroscopy in Astrophysics. pp
  297--298, \mn@doi{10.1007/978-3-540-75485-5_72}

\bibitem[\protect\citeauthoryear{{Mayor} et~al.,}{{Mayor}
  et~al.}{2003}]{2003Msngr.114...20M}
{Mayor} M.,  et~al., 2003, The Messenger, \href
  {http://adsabs.harvard.edu/abs/2003Msngr.114...20M} {114, 20}

\bibitem[\protect\citeauthoryear{{Molaro} \& {Centuri{\'o}n}}{{Molaro} \&
  {Centuri{\'o}n}}{2011}]{2011A&A...525A..74M}
{Molaro} P.,  {Centuri{\'o}n} M.,  2011, \mn@doi [\aap]
  {10.1051/0004-6361/201015179}, \href
  {http://adsabs.harvard.edu/abs/2011A%26A...525A..74M} {525, A74}

\bibitem[\protect\citeauthoryear{{Molaro}, {Reimers}, {Agafonova}  \&
  {Levshakov}}{{Molaro} et~al.}{2008a}]{2008EPJST.163..173M}
{Molaro} P.,  {Reimers} D.,  {Agafonova} I.~I.,   {Levshakov} S.~A.,  2008a,
  \mn@doi [European Physical Journal Special Topics]
  {10.1140/epjst/e2008-00818-4}, \href
  {http://adsabs.harvard.edu/abs/2008EPJST.163..173M} {163, 173}

\bibitem[\protect\citeauthoryear{{Molaro}, {Levshakov}, {Monai},
  {Centuri{\'o}n}, {Bonifacio}, {D'Odorico}  \& {Monaco}}{{Molaro}
  et~al.}{2008b}]{2008A&A...481..559M}
{Molaro} P.,  {Levshakov} S.~A.,  {Monai} S.,  {Centuri{\'o}n} M.,  {Bonifacio}
  P.,  {D'Odorico} S.,   {Monaco} L.,  2008b, \mn@doi [\aap]
  {10.1051/0004-6361:20078864}, \href
  {http://adsabs.harvard.edu/abs/2008A%26A...481..559M} {481, 559}

\bibitem[\protect\citeauthoryear{{Molaro} et~al.,}{{Molaro}
  et~al.}{2013a}]{2013A&A...555A..68M}
{Molaro} P.,  et~al., 2013a, \mn@doi [\aap] {10.1051/0004-6361/201321351},
  \href {http://adsabs.harvard.edu/abs/2013A%26A...555A..68M} {555, A68}

\bibitem[\protect\citeauthoryear{{Molaro} et~al.,}{{Molaro}
  et~al.}{2013b}]{2013A&A...560A..61M}
{Molaro} P.,  et~al., 2013b, \mn@doi [\aap] {10.1051/0004-6361/201322324},
  \href {http://adsabs.harvard.edu/abs/2013A%26A...560A..61M} {560, A61}

\bibitem[\protect\citeauthoryear{{Murphy}}{{Murphy}}{2016}]{michael_murphy_2016_44765}
{Murphy} M.~T.,  2016, {UVES\_popler: POst-PipeLine Echelle Reduction
  software}, \mn@doi{10.5281/zenodo.44765}, \url
  {http://dx.doi.org/10.5281/zenodo.44765}

\bibitem[\protect\citeauthoryear{{Murphy} \& {Berengut}}{{Murphy} \&
  {Berengut}}{2014}]{2014MNRAS.438..388M}
{Murphy} M.~T.,  {Berengut} J.~C.,  2014, \mn@doi [\mnras]
  {10.1093/mnras/stt2204}, \href
  {http://adsabs.harvard.edu/abs/2014MNRAS.438..388M} {438, 388}

\bibitem[\protect\citeauthoryear{{Murphy}, {Webb}, {Flambaum}, {Dzuba},
  {Churchill}, {Prochaska}, {Barrow}  \& {Wolfe}}{{Murphy}
  et~al.}{2001a}]{2001MNRAS.327.1208M}
{Murphy} M.~T.,  {Webb} J.~K.,  {Flambaum} V.~V.,  {Dzuba} V.~A.,  {Churchill}
  C.~W.,  {Prochaska} J.~X.,  {Barrow} J.~D.,   {Wolfe} A.~M.,  2001a, \mn@doi
  [\mnras] {10.1046/j.1365-8711.2001.04840.x}, \href
  {http://adsabs.harvard.edu/abs/2001MNRAS.327.1208M} {327, 1208}

\bibitem[\protect\citeauthoryear{{Murphy}, {Webb}, {Flambaum}, {Churchill}  \&
  {Prochaska}}{{Murphy} et~al.}{2001b}]{2001MNRAS.327.1223M}
{Murphy} M.~T.,  {Webb} J.~K.,  {Flambaum} V.~V.,  {Churchill} C.~W.,
  {Prochaska} J.~X.,  2001b, \mn@doi [\mnras]
  {10.1046/j.1365-8711.2001.04841.x}, \href
  {http://adsabs.harvard.edu/abs/2001MNRAS.327.1223M} {327, 1223}

\bibitem[\protect\citeauthoryear{{Murphy}, {Webb}, {Flambaum}  \&
  {Curran}}{{Murphy} et~al.}{2003a}]{2003Ap&SS.283..577M}
{Murphy} M.~T.,  {Webb} J.~K.,  {Flambaum} V.~V.,   {Curran} S.~J.,  2003a,
  \mn@doi [\apss] {10.1023/A:1022570532369}, \href
  {http://adsabs.harvard.edu/abs/2003Ap%26SS.283..577M} {283, 577}

\bibitem[\protect\citeauthoryear{{Murphy}, {Webb}  \& {Flambaum}}{{Murphy}
  et~al.}{2003b}]{2003MNRAS.345..609M}
{Murphy} M.~T.,  {Webb} J.~K.,   {Flambaum} V.~V.,  2003b, \mn@doi [\mnras]
  {10.1046/j.1365-8711.2003.06970.x}, \href
  {http://adsabs.harvard.edu/abs/2003MNRAS.345..609M} {345, 609}

\bibitem[\protect\citeauthoryear{{Murphy}, {Flambaum}, {Webb}, {Dzuba},
  {Prochaska}  \& {Wolfe}}{{Murphy} et~al.}{2004}]{2004LNP...648..131M}
{Murphy} M.~T.,  {Flambaum} V.~V.,  {Webb} J.~K.,  {Dzuba} V.~A.,  {Prochaska}
  J.~X.,   {Wolfe} A.~M.,  2004, in {Karshenboim} S.~G.,  {Peik} E.,  eds,
  Lecture Notes in Physics, Berlin Springer Verlag Vol. 648, Astrophysics,
  Clocks and Fundamental Constants. pp 131--150 (\mn@eprint {}
  {astro-ph/0310318})

\bibitem[\protect\citeauthoryear{{Murphy}, {Webb}  \& {Flambaum}}{{Murphy}
  et~al.}{2007a}]{2007PhRvL..99w9001M}
{Murphy} M.~T.,  {Webb} J.~K.,   {Flambaum} V.~V.,  2007a, \mn@doi [Physical
  Review Letters] {10.1103/PhysRevLett.99.239001}, \href
  {http://adsabs.harvard.edu/abs/2007PhRvL..99w9001M} {99, 239001}

\bibitem[\protect\citeauthoryear{{Murphy}, {Tzanavaris}, {Webb}  \&
  {Lovis}}{{Murphy} et~al.}{2007b}]{2007MNRAS.378..221M}
{Murphy} M.~T.,  {Tzanavaris} P.,  {Webb} J.~K.,   {Lovis} C.,  2007b, \mn@doi
  [\mnras] {10.1111/j.1365-2966.2007.11768.x}, \href
  {http://adsabs.harvard.edu/abs/2007MNRAS.378..221M} {378, 221}

\bibitem[\protect\citeauthoryear{{Murphy} et~al.,}{{Murphy}
  et~al.}{2007c}]{2007MNRAS.380..839M}
{Murphy} M.~T.,  et~al., 2007c, \mn@doi [\mnras]
  {10.1111/j.1365-2966.2007.12147.x}, \href
  {http://adsabs.harvard.edu/abs/2007MNRAS.380..839M} {380, 839}

\bibitem[\protect\citeauthoryear{{Murphy}, {Flambaum}, {Muller}  \&
  {Henkel}}{{Murphy} et~al.}{2008a}]{2008Sci...320.1611M}
{Murphy} M.~T.,  {Flambaum} V.~V.,  {Muller} S.,   {Henkel} C.,  2008a, \mn@doi
  [Science] {10.1126/science.1156352}, \href
  {http://adsabs.harvard.edu/abs/2008Sci...320.1611M} {320, 1611}

\bibitem[\protect\citeauthoryear{{Murphy}, {Webb}  \& {Flambaum}}{{Murphy}
  et~al.}{2008b}]{2008MNRAS.384.1053M}
{Murphy} M.~T.,  {Webb} J.~K.,   {Flambaum} V.~V.,  2008b, \mn@doi [\mnras]
  {10.1111/j.1365-2966.2007.12695.x}, \href
  {http://adsabs.harvard.edu/abs/2008MNRAS.384.1053M} {384, 1053}

\bibitem[\protect\citeauthoryear{{Pepe} et~al.,}{{Pepe}
  et~al.}{2010}]{2010SPIE.7735E..0FP}
{Pepe} F.~A.,  et~al., 2010, in Society of Photo-Optical Instrumentation
  Engineers (SPIE) Conference Series. p. 77350F, \mn@doi{10.1117/12.857122}

\bibitem[\protect\citeauthoryear{{Piskunov} \& {Valenti}}{{Piskunov} \&
  {Valenti}}{2002}]{2002A&A...385.1095P}
{Piskunov} N.~E.,  {Valenti} J.~A.,  2002, \mn@doi [\aap]
  {10.1051/0004-6361:20020175}, \href
  {http://adsabs.harvard.edu/abs/2002A%26A...385.1095P} {385, 1095}

\bibitem[\protect\citeauthoryear{{Quast}, {Reimers}  \& {Levshakov}}{{Quast}
  et~al.}{2004}]{2004A&A...415L...7Q}
{Quast} R.,  {Reimers} D.,   {Levshakov} S.~A.,  2004, \mn@doi [\aap]
  {10.1051/0004-6361:20040013}, \href
  {http://adsabs.harvard.edu/abs/2004A%26A...415L...7Q} {415, L7}

\bibitem[\protect\citeauthoryear{{Rahmani} et~al.,}{{Rahmani}
  et~al.}{2013}]{2013MNRAS.435..861R}
{Rahmani} H.,  et~al., 2013, \mn@doi [\mnras] {10.1093/mnras/stt1356}, \href
  {http://adsabs.harvard.edu/abs/2013MNRAS.435..861R} {435, 861}

\bibitem[\protect\citeauthoryear{{Reimers}, {Hagen}, {Rodriguez-Pascual}  \&
  {Wisotzki}}{{Reimers} et~al.}{1998}]{1998A&A...334...96R}
{Reimers} D.,  {Hagen} H.-J.,  {Rodriguez-Pascual} P.,   {Wisotzki} L.,  1998,
  \aap, \href {http://adsabs.harvard.edu/abs/1998A%26A...334...96R} {334, 96}

\bibitem[\protect\citeauthoryear{{Srianand}, {Chand}, {Petitjean}  \&
  {Aracil}}{{Srianand} et~al.}{2004}]{2004PhRvL..92l1302S}
{Srianand} R.,  {Chand} H.,  {Petitjean} P.,   {Aracil} B.,  2004, \mn@doi
  [Physical Review Letters] {10.1103/PhysRevLett.92.121302}, \href
  {http://adsabs.harvard.edu/abs/2004PhRvL..92l1302S} {92, 121302}

\bibitem[\protect\citeauthoryear{{Srianand}, {Chand}, {Petitjean}  \&
  {Aracil}}{{Srianand} et~al.}{2007}]{2007PhRvL..99w9002S}
{Srianand} R.,  {Chand} H.,  {Petitjean} P.,   {Aracil} B.,  2007, \mn@doi
  [Physical Review Letters] {10.1103/PhysRevLett.99.239002}, \href
  {http://adsabs.harvard.edu/abs/2007PhRvL..99w9002S} {99, 239002}

\bibitem[\protect\citeauthoryear{{Szentgyorgyi} et~al.,}{{Szentgyorgyi}
  et~al.}{2014}]{2014SPIE.9147E..26S}
{Szentgyorgyi} A.,  et~al., 2014, in Ground-based and Airborne Instrumentation
  for Astronomy V. p. 914726, \mn@doi{10.1117/12.2056741}

\bibitem[\protect\citeauthoryear{{Uzan}}{{Uzan}}{2011}]{2011LRR....14....2U}
{Uzan} J.-P.,  2011, \mn@doi [Living Reviews in Relativity]
  {10.12942/lrr-2011-2}, \href
  {http://adsabs.harvard.edu/abs/2011LRR....14....2U} {14, 2}

\bibitem[\protect\citeauthoryear{{Webb}, {Flambaum}, {Churchill}, {Drinkwater}
  \& {Barrow}}{{Webb} et~al.}{1999}]{1999PhRvL..82..884W}
{Webb} J.~K.,  {Flambaum} V.~V.,  {Churchill} C.~W.,  {Drinkwater} M.~J.,
  {Barrow} J.~D.,  1999, \mn@doi [Physical Review Letters]
  {10.1103/PhysRevLett.82.884}, \href
  {http://adsabs.harvard.edu/abs/1999PhRvL..82..884W} {82, 884}

\bibitem[\protect\citeauthoryear{{Webb}, {Murphy}, {Flambaum}, {Dzuba},
  {Barrow}, {Churchill}, {Prochaska}  \& {Wolfe}}{{Webb}
  et~al.}{2001}]{2001PhRvL..87i1301W}
{Webb} J.~K.,  {Murphy} M.~T.,  {Flambaum} V.~V.,  {Dzuba} V.~A.,  {Barrow}
  J.~D.,  {Churchill} C.~W.,  {Prochaska} J.~X.,   {Wolfe} A.~M.,  2001,
  \mn@doi [Physical Review Letters] {10.1103/PhysRevLett.87.091301}, \href
  {http://adsabs.harvard.edu/abs/2001PhRvL..87i1301W} {87, 091301}

\bibitem[\protect\citeauthoryear{{Webb}, {King}, {Murphy}, {Flambaum},
  {Carswell}  \& {Bainbridge}}{{Webb} et~al.}{2011}]{2011PhRvL.107s1101W}
{Webb} J.~K.,  {King} J.~A.,  {Murphy} M.~T.,  {Flambaum} V.~V.,  {Carswell}
  R.~F.,   {Bainbridge} M.~B.,  2011, \mn@doi [Physical Review Letters]
  {10.1103/PhysRevLett.107.191101}, \href
  {http://adsabs.harvard.edu/abs/2011PhRvL.107s1101W} {107, 191101}

\bibitem[\protect\citeauthoryear{{Whitmore} \& {Murphy}}{{Whitmore} \&
  {Murphy}}{2015}]{2015MNRAS.447..446W}
{Whitmore} J.~B.,  {Murphy} M.~T.,  2015, \mn@doi [\mnras]
  {10.1093/mnras/stu2420}, \href
  {http://adsabs.harvard.edu/abs/2015MNRAS.447..446W} {447, 446}

\bibitem[\protect\citeauthoryear{{Whitmore}, {Murphy}  \& {Griest}}{{Whitmore}
  et~al.}{2010}]{2010ApJ...723...89W}
{Whitmore} J.~B.,  {Murphy} M.~T.,   {Griest} K.,  2010, \mn@doi [\apj]
  {10.1088/0004-637X/723/1/89}, \href
  {http://adsabs.harvard.edu/abs/2010ApJ...723...89W} {723, 89}

\bibitem[\protect\citeauthoryear{{Wilczynska}, {Webb}, {King}, {Murphy},
  {Bainbridge}  \& {Flambaum}}{{Wilczynska} et~al.}{2015}]{2015MNRAS.454.3082W}
{Wilczynska} M.~R.,  {Webb} J.~K.,  {King} J.~A.,  {Murphy} M.~T.,
  {Bainbridge} M.~B.,   {Flambaum} V.~V.,  2015, \mn@doi [\mnras]
  {10.1093/mnras/stv2148}, \href
  {http://adsabs.harvard.edu/abs/2015MNRAS.454.3082W} {454, 3082}

\bibitem[\protect\citeauthoryear{{Wilken} et~al.,}{{Wilken}
  et~al.}{2010}]{2010MNRAS.405L..16W}
{Wilken} T.,  et~al., 2010, \mn@doi [\mnras]
  {10.1111/j.1745-3933.2010.00850.x}, \href
  {http://adsabs.harvard.edu/abs/2010MNRAS.405L..16W} {405, L16}

\bibitem[\protect\citeauthoryear{{Zerbi} et~al.,}{{Zerbi}
  et~al.}{2014}]{2014SPIE.9147E..23Z}
{Zerbi} F.~M.,  et~al., 2014, in Ground-based and Airborne Instrumentation for
  Astronomy V. p. 914723, \mn@doi{10.1117/12.2055329}

\makeatother
\end{thebibliography}

\bsp

\label{lastpage}

\end{document}